\documentclass[prd,preprint,superscriptaddress,showpacs,nofootinbib,tightenlines,amsmath]{revtex4-1}
\usepackage{graphicx}
\usepackage{epsfig}
\usepackage{slashed}
\usepackage{braket}
\usepackage{nicefrac}
\usepackage{bbm}
\usepackage{amssymb}
\usepackage{color}
\usepackage{bbold}

\def\no{\nonumber}
 \def\la{\langle}
 \def\ra{\rangle}

 \def\chip{\chi_+}
 
 \def\chim{\chi_-}

\newcommand{\dega}{\dagger}

\newcommand{\br}{\langle}
\newcommand{\ke}{\rangle}

\newcommand{\bda}{\begin{\displaymath}\begin{array}{rl}}
\newcommand{\eda}{\end{array}\end{displaymath}}
\newcommand{\be}{\begin{equation}}
\newcommand{\ee}{\end{equation}}
\newcommand{\bdm}{\begin{displaymath}}
\newcommand{\edm}{\end{displaymath}}
\newcommand{\bea}{\begin{eqnarray}}
\newcommand{\eea}{\end{eqnarray}}

\newcommand{\myTR}{\hline\hline}
\newcommand{\myMR}{\hline}
\newcommand{\myBR}{\hline\hline}

\begin{document}
\title{Two-photon decays and transition form factors of $\pi^0$, $\eta$, and $\eta'$\\ in large-$N_c$ chiral perturbation theory}
\author{P.~Bickert}
\affiliation{Institut f\"ur
Kernphysik, Johannes Gutenberg-Universit\"at Mainz, D-55099 Mainz,
Germany}
\affiliation{Fraunhofer-Institut f\"ur Techno- und Wirtschaftsmathematik ITWM, D-67663 Kaiserslautern, Germany}
\author{S.~Scherer}
\affiliation{Institut f\"ur
Kernphysik, Johannes Gutenberg-Universit\"at Mainz, D-55099 Mainz,
Germany}
\date{\today}
%\preprint{MITP/}
\begin{abstract}
	We present a calculation of $P\to \gamma^{(\ast)}\gamma^{(\ast)}$ processes, where $P=\pi^0,\ \eta,\ \eta'$, at the one-loop level up to and including next-to-next-to-leading order (NNLO) in large-$N_c$ chiral perturbation theory.
	The results are numerically evaluated successively at LO, NLO, and NNLO. The appearing low-energy	constants are determined through fits to the available experimental data. We investigate the decay widths to real photons, the single-virtual transition form factors, and the widths of $P\to\gamma l^+l^-$, where $l=e,\ \mu$. Furthermore, we provide results for the slopes and curvatures of the transition form factors.
\end{abstract}
%\pacs{}
\maketitle

\section{Introduction}

   In recent years, the two-photon interaction of the light pseudoscalar mesons 
has received considerable attention from both the experimental and theoretical sides
\cite{Danilkin:2019mhd}.
To a large extent, this renewed interest was triggered by the muon anomalous
magnetic moment discrepancy which states a 3.3 sigma deviation between experiment and theory (see Refs.~\cite{Danilkin:2019mhd,Jegerlehner:2009ry,Hoecker:2019} for  reviews).
On the theoretical side, the largest uncertainty in the anomalous magnetic 
moment $a_\mu$ originates from the evaluation of hadronic contributions, namely, the hadronic vacuum polarization and the hadronic light-by-light (HLbL) scattering \cite{Hoecker:2019}.
In this context, the two-photon decays of the light pseudoscalars enter the HLbL
contribution in terms of pseudoscalar-exchange diagrams (see Fig.~35 of Ref.\ \cite{Jegerlehner:2009ry}).

Besides this more phenomenology-driven interest, the two-photon decays of the
light pseudoscalars provide an ideal laboratory for investigating the symmetry-breaking mechanisms relevant in quantum chromodynamics (QCD).
To be specific, the low-energy regime of QCD is characterized by an interplay
between the dynamical (spontaneous) breaking of chiral symmetry, the explicit symmetry breaking by the quark masses, and the axial $\text{U(1)}_A$ anomaly.
It is a generally accepted feature of QCD that the global
$\text{U(3)}_L\times\text{U(3)}_R$ chiral symmetry of the QCD Lagrangian at the classical level for vanishing up-, down-, and strange-quark masses is dynamically broken down to $\mbox{SU(3)}_V\times\mbox{U(1)}_V$ in the ground state (see, e.g.,
Ref.~\cite{Scherer:2012zzd} for a discussion).
Naively, one would then expect the appearance of nine massless pseudoscalar 
Goldstone bosons \cite{Goldstone:1962es}.
However, because of quantum effects, the singlet axial-vector current is no longer 
conserved (U(1)$_A$ anomaly) and the corresponding alleged singlet Goldstone boson acquires a mass even in the chiral limit of massless quarks
\cite{'tHooft:1976up,Witten:1979vv,Veneziano:1979ec}.
At this stage, the large-number-of-colors (L$N_c$) limit of 
QCD~\cite{'tHooft:1973jz,Witten:1979kh}, i.e., $N_c\to\infty$ with $g^2N_c$ fixed,
provides another theoretical simplification aside from the assumption of massless quarks.
Since the four divergence of the anomalous singlet axial-vector current is
proportional to the square of the strong coupling constant $g$ \cite{Weinberg:1975ui}, it vanishes in the L$N_c$ limit.
Therefore, the singlet pseudoscalar is also a Goldstone boson in the
combined chiral and L$N_c$ limits, resulting in total in a pseudoscalar nonet
$(\pi,K,\eta_8,\eta_1)$ as the Goldstone bosons \cite{Witten:1979vv,Coleman:1980mx}.
Of course, massless L$N_c$ QCD is only an approximation to the real world.
However, one may use it as a starting point for a perturbative framework, treating 
the symmetry breaking by the U(1)$_A$ anomaly and by the nonzero quark masses as corrections.

At leading order in the $1/N_c$ and quark-mass expansion, the decays
$P\to\gamma\gamma$ $(P=\pi^0,\eta,\eta')$ are driven by the chiral anomaly in terms of the Wess-Zumino-Witten effective action \cite{Wess:1971yu,Witten:1983tw} (see, e.g., Ref.~\cite{Scherer:2012zzd} for an introduction).
Corrections to the WZW predictions originate from the axial U(1)$_A$ anomaly and 
the nonzero quark masses.
Both mechanisms are also reponsible for generating the masses of the originally 
massless Goldstone bosons and for the $\eta$-$\eta'$ mixing.  
These modifications may be systematically calculated in the framework of 
large-$N_c$ chiral perturbation theory (L$N_c$ChPT) \cite{Leutwyler:1996sa,HerreraSiklody:1996pm,Kaiser:2000gs},
which can be viewed as an extension of conventional ChPT \cite{Gasser:1984gg} by including, in addition to the pseudoscalar octet, the pseudoscalar singlet.
In L$N_c$ChPT, the most general effective Lagrangian is organized in a combined 
expansion in terms of momenta (derivatives), quark masses, and $1/N_c$.
Observables are calculated perturbatively, according to a power counting with 
respect to a collective small expansion parameter $\delta$ \cite{Leutwyler:1996sa}.

In this article, we investigate the $P\to \gamma^{(\ast)}\gamma^{(\ast)}$ 
interaction at next-to-next-to-leading order (NNLO) in L$N_c$ChPT.
In Sec.~{II}, we describe the effective field theory 
we will consider for our calculation by specifying the Lagrangian and the power counting. 
In Sec.~{III}, we define the invariant amplitude and discuss its perturbative 
calculation including the $\eta$-$\eta'$ mixing. 
Section~IV contains our numerical results for the decay rates and the transition
form factors (TFF) at LO, NLO, and NNLO, respectively.
In Sec.~V, we discuss the decay rates for single Dalitz decays. 
Finally, in Sec.~VI we conclude with a few remarks and an outlook on possible 
future work.

\section{Lagrangians and power counting}
\label{sec:Lagrangians}
 In the framework of L$N_c$ChPT, one performs a simultaneous expansion of (renormalized) Feynman diagrams
in terms of momenta $p$, quark masses $m$, and $1/N_c$.\footnote{It
is understood that dimensionful variables need to be small in
comparison with an energy scale.}
   Introducing a collective expansion parameter $\delta$, the variables are counted as small quantities of order
\cite{Leutwyler:1996sa}
\begin{equation}
\label{powerexp}
p=\mathcal{O}(\sqrt{\delta}),\ \ \ m=\mathcal{O}(\delta),\ \ \
1/N_c=\mathcal{O}(\delta).
\end{equation}
   The most general Lagrangian of L$N_c$ChPT is organized as an
infinite series in terms of derivatives, quark-mass terms, and,
implicitly, powers of $1/N_c$, with the scaling behavior given in
Eq.~(\ref{powerexp}):
\begin{equation}
\label{Leff}
\mathcal{L}_{\text{eff}}=\mathcal{L}^{(0)}+\mathcal{L}^{(1)}+\mathcal{L}^{(2)}+\mathcal{L}^{(3)}+\dots,
\end{equation}
   where the superscripts $(i)$ denote the order in $\delta$.

   The dynamical degrees of freedom are collected in the unitary $3\times 3$ matrix
\begin{equation}
\label{definitionU}
 U(x)=\exp\left(i\frac{\phi(x)}{F}\right),
\end{equation}
where the Hermitian $3\times 3$ matrix
\begin{align}
\label{eq:pseudoscalar_mesons}
\phi=\sum_{a=0}^8\phi_a\lambda_a=
\begin{pmatrix}
\pi^0+\frac{1}{\sqrt{3}}\eta_8+\sqrt{\frac{2}{3}}\eta_1 & \sqrt{2}\pi^+ & \sqrt{2}K^+\\
\sqrt{2}\pi^- &  -\pi^0+\frac{1}{\sqrt{3}}\eta_8+\sqrt{\frac{2}{3}}\eta_1 & \sqrt{2}K^0\\
\sqrt{2}K^- & \sqrt{2}\bar{K}^0 & -\frac{2}{\sqrt{3}}\eta_8+\sqrt{\frac{2}{3}}\eta_1\\
\end{pmatrix}
\end{align}
contains the pseudoscalar octet fields and the pseudoscalar singlet field $\eta_1$, the
$\lambda_a$ ($a=1,\ldots,8$) are the Gell-Mann matrices, and
$\lambda_0=\sqrt{2/3}\, {\mathbbm 1}$.
   In Eq.~(\ref{definitionU}), $F$ denotes the pion-decay constant in the three-flavor chiral limit and is counted as $F=\mathcal{O}(\sqrt{N_c})=\mathcal{O}(1/\sqrt{\delta})$\footnote{Here, we deviate from the often-used convention of indicating the {\it three}-flavor chiral limit by a subscript 0.} \cite{Witten:1979vv}.
   In addition to the dynamical degrees of freedom of Eq.~(\ref{eq:pseudoscalar_mesons}),
the effective Lagrangian also contains a set of external fields $(s,p,l_\mu,r_\mu,\theta)$.
   The fields $s$, $p$, $l_\mu$, and $r_\mu$ are Hermitian, color-neutral $3\times 3$ matrices
coupling to the corresponding quark bilinears, and $\theta$ is a real field
coupling to the winding number density \cite{Gasser:1984gg}.
   The external scalar and pseudoscalar fields $s$ and $p$ are combined
in the definition $\chi\equiv 2B(s+ip)$ \cite{Gasser:1984gg}.
   The low-energy constant (LEC) $B$ is related to the scalar singlet quark condensate
$\langle\bar{q}q\rangle_0$ in the three-flavor chiral limit and is of
${\cal O}(N_c^0)$ \cite{Leutwyler:1996sa}.

In general, applying the power counting of Eq.~(\ref{powerexp}) to the construction of the effective
Lagrangian in the L$N_c$ framework involves two ingredients.
   On the one hand, there is the momentum and quark-mass counting which proceeds as in conventional $\text{SU}(3)$ ChPT
\cite{Gasser:1984gg}: (covariant) derivatives count as ${\cal O}(p)$, $\chi$ counts as ${\cal O}(p^2)$, etc. We denote the corresponding chiral order by $D_p$.
   The L$N_c$ behavior can be determined by using the following rules (see Refs.~\cite{HerreraSiklody:1996pm,Kaiser:2000gs}
for a detailed account).
   In the L$N_c$ counting, the leading contribution to a quark correlation function is given by a single flavor trace and is of order $N_c$ \cite{'tHooft:1973jz,Witten:1979kh,Manohar:1998xv}.
   In general, diagrams with $r$ quark loops and thus $r$ flavor traces are of order $N_c^{2-r}$. Terms
without traces correspond to the purely gluonic theory and count at leading order as $N_c^2$.
   This argument is transferred to the level of the effective Lagrangian, i.e., single-trace terms are
of order $N_c$, double-trace terms of order unity, etc.\footnote{When applying these counting rules,
one has to account for the so-called trace relations connecting single-trace terms with products of traces
(see, e.g., Appendix A of Ref.~\cite{Fearing:1994ga}).}
  
   Introducing $\psi=\sqrt{6} \eta_1/F$ \cite{Kaiser:2000gs},
	 each power $(\psi+\theta)^n$ is accompanied by a coefficient of order ${\cal O}(N_c^{-n})$.
   The reason for this assignment is the fact that, in QCD, the external field $\theta$ couples
to the winding number density with strength $1/N_c$.
   In a similar fashion, $D_\mu\theta$ (as well as multiple derivatives) are related to expressions
with ${\cal O}(N_c^{-1})$.\footnote{Note that we do not directly book the quantities $(\psi+\theta)$
or $D_\mu\theta$ as ${\cal O}(N_c^{-1})$, but rather attribute this order to the coefficients
coming with the terms.}
   Denoting the number of $(\psi+\theta)$ and $D_\mu\theta$ terms by $N_\theta$,
the L$N_c$ order reads \cite{HerreraSiklody:1996pm,Kaiser:2000gs}
\begin{equation}
\label{DNc}
D_{N^{-1}_c}=-2+N_{tr}+N_\theta.
\end{equation}
   The combined order of an operator is then given by
\begin{equation}
\label{Ddelta}
D_\delta=\frac{1}{2}D_p+D_{N_c^{-1}}.
\end{equation}

\subsection{Wess-Zumino-Witten effective action}

The two-photon decays arise from the odd-intrinsic-parity part of the effective field theory. At leading order, they are driven by the chiral anomaly, which is accounted for by the Wess-Zumino-Witten (WZW) action \cite{Wess:1971yu,Witten:1983tw}. In U(3) ChPT, the WZW action (without external fields) reads
\begin{align}
\label{WZWaction}
S^0_{\text{ano}}&=N_c\,S^0_{\text{WZW}},\no\\
S^0_{\text{WZW}}&=-\frac{i}{240\pi^2}\int^1_0 d\alpha\int d^4x\epsilon^{ijklm}\la\mathcal{U}^L_i \mathcal{U}^L_j \mathcal{U}^L_k \mathcal{U}^L_l \mathcal{U}^L_m\ra,
\end{align}
where $\la \dots \ra$ denotes the (flavor) trace.
For the construction of the WZW action, the domain of definition of $U$ needs to be extended to a (hypothetical) fifth dimension,
\begin{align}
U(y)=\exp\left(i\alpha\frac{\phi(x)}{F}\right),\ \ y^i=(x^\mu,\alpha),\ i=0,\dots,4,\ 0\leq\alpha\leq 1,
\end{align}
where Minkowski space is defined as the surface of the five-dimensional space for $\alpha=1$. The indices $i,\dots,m$ in Eq.~(\ref{WZWaction}) run from 0 to 4, $y_4=y^4=\alpha$, $\epsilon_{ijklm}$ is the completely antisymmetric (five-dimensional) tensor with $\epsilon_{01234}=-\epsilon^{01234}=1$, and $\mathcal{U}^L_i=U^\dagger \partial U/\partial y^i$. 
		In the presence of external fields, the anomalous action receives an additional term \cite{Manes:1984gk,Bijnens:1993xi}
\begin{align}
S_{\text{ano}}=N_c(S^0_{\text{WZW}}+S^{\text{ext}}_{\text{WZW}})
\label{WZWextfields}
\end{align}
given by
\begin{align}
S^{\text{ext}}_{\text{WZW}}=-\frac{i}{48\pi^2}\int d^4x \epsilon^{\mu\nu\rho\sigma}\left\{\la Z_{\mu\nu\rho\sigma}(U,l,r)\ra-\la Z_{\mu\nu\rho\sigma}(\mathbb{1},l,r)\ra\right\},
\end{align}
with
\begin{align}
&Z_{\mu\nu\rho\sigma}(U,l,r)\no\\
&=\frac{1}{2}U l_\mu U^\dagger r_\nu U l_\rho U^\dagger r_\sigma+U l_\mu l_\nu l_\rho U^\dagger r_\sigma-U^\dagger r_\mu r_\nu r_\rho U l_\sigma\no\\
&\quad + i U \partial_\mu l_\nu l_\rho U^\dagger r_\sigma-i U^\dagger \partial_\mu r_\nu r_\rho U l_\sigma+i\partial_\mu r_\nu U l_\rho U^\dagger r_\sigma-i\partial_\mu l_\nu U^\dagger r_\rho U^\dagger l_\sigma\no\\
&\quad - i \mathcal{U}_{L\mu}  l_\nu U^\dagger r_\rho U l_\sigma+i \mathcal{U}_{R\mu} r_\nu U l_\rho U^\dagger r_\sigma-i\mathcal{U}_{L\mu} l_\nu l_\rho l_\sigma+i\mathcal{U}_{R\mu} r_\nu r_\rho r_\sigma\no\\
&\quad +\frac{1}{2}\left(\mathcal{U}_{L\mu} U^\dagger \partial_\nu  r_\rho U l_\sigma- \mathcal{U}_{R\mu}U \partial_\nu l_\rho U^\dagger r_\sigma+\mathcal{U}_{L\mu}U^\dagger r_\nu U \partial_\rho l_\sigma-\mathcal{U}_{R\mu} U l_\nu U^\dagger \partial_\rho r_\sigma\right)\no\\
&\quad -\mathcal{U}_{L\mu} \mathcal{U}_{L\nu} U^\dagger r_\rho U l_\sigma+ \mathcal{U}_{R\mu} \mathcal{U}_{R\nu}U l_\rho U^\dagger r_\sigma+\frac{1}{2}\mathcal{U}_{L\mu} l_\nu U \mathcal{U}_{L\rho} l_\sigma-\frac{1}{2}\mathcal{U}_{R\mu} r_\nu \mathcal{U}_{R\rho} r_\sigma\no\\
&\quad +\mathcal{U}_{L\mu} l_\nu \partial_\rho l_\sigma - \mathcal{U}_{R\mu} r_\nu \partial_\rho r_\sigma+\mathcal{U}_{L\mu} \partial_\nu l_\rho l_\sigma-\mathcal{U}_{R\mu} \partial_\nu r_\rho r_\sigma\no\\
&\quad -i\mathcal{U}_{L\mu} \mathcal{U}_{L\nu} \mathcal{U}_{L\rho} l_\sigma+i \mathcal{U}_{R\mu} \mathcal{U}_{R\nu} \mathcal{U}_{R\rho} r_\sigma,
\end{align}
where $\mathcal{U}_{L\mu}\equiv U^\dagger\partial_\mu U$ and $\mathcal{U}_{R\mu}\equiv U\partial_\mu U^\dagger$. The subtraction of the $\la Z_{\mu\nu\rho\sigma}(\mathbb{1},l,r)\ra$ term is necessary to satisfy a boundary condition leading to an action that is consistent with the conservation of the vector current.

\subsection{Normal-parity Lagrangians}

In the NNLO calculation of the two-photon decays, the LO, NLO, and NNLO Lagrangians of even intrinsic parity enter as well. The leading-order Lagrangian is given by \cite{Leutwyler:1996sa,Kaiser:2000gs}
\begin{equation}
\label{eq:lolagrangian}
\mathcal{L}^{(0)}=\frac{F^2}{4}\langle D_\mu U D^\mu U^\dagger\rangle+\frac{F^2}{4}\langle\chi U^\dagger+U\chi^\dagger\rangle-\frac{1}{2}\tau(\psi+\theta)^2,
\end{equation}
where the covariant derivatives of $U$ and $U^\dagger$ are defined as
\begin{align}
D_\mu U&=\partial_\mu U-ir_\mu U+i U l_\mu,\no\\
D_\mu U^\dagger&=\partial_\mu U^\dagger+iU^\dagger r_\mu -i l_\mu U^\dagger.
\end{align}
The constant $\tau=\mathcal{O}(N_c^0)$ is the topological
susceptibility of the purely gluonic theory \cite{Leutwyler:1996sa}.
Counting the quark mass as ${\cal O}(p^2)$, the first two terms of $\mathcal{L}^{(0)}$
are of $\mathcal{O}(N_c p^2)$, while the third term is of $\mathcal{O}(N_c^0)$, i.e.,~all terms are of ${\cal O}(\delta^{0})$.

The normal-parity part of the NLO Lagrangian $\mathcal{L}^{(1)}$ was constructed in Refs.~\cite{Leutwyler:1996sa,HerreraSiklody:1996pm,Kaiser:2000gs}
and receives contributions of $\mathcal{O}(N_c p^4)$, $\mathcal{O}(p^2)$, and $\mathcal{O}(N_c^{-1})$.
We only display the terms relevant for our calculation, in particular, here, we set
$v_\mu\equiv (r_\mu+l_\mu)/2=0$ and $a_\mu\equiv(r_\mu-l_\mu)/2=0$ in the covariant derivatives:
\begin{align}
	\label{L1}
	\mathcal{L}^{(1)}&= L_5 \langle D_\mu U D^\mu U^\dagger (\chi U^\dagger + U \chi^\dagger)\rangle
	+ L_8 \langle \chi U^\dagger\chi U^\dagger + U \chi^\dagger U \chi^\dagger \rangle  \no \\
	&\quad + \frac{F^2}{12} \Lambda_1 D_\mu \psi D^\mu \psi - i \frac{F^2}{12} \Lambda_2 (\psi+\theta)
	\langle \chi U^\dagger - U \chi^\dagger \rangle+\dots,
\end{align}
where
\begin{align}
	\label{Dmupsi}
	D_\mu\psi&=\partial_\mu\psi-2\langle a_\mu\rangle,
\end{align}
and the ellipsis refers to the neglected terms.
The first two terms of $\mathcal{L}^{(1)}$ count as $\mathcal{O}(N_cp^4)$ and are obtained from the standard
SU(3) ChPT Lagrangian of ${\cal O}(p^4)$ \cite{Gasser:1984gg} by retaining solely terms with a single trace
and keeping only the constant terms of the so-called potentials which are functions of $\psi+\theta$ \cite{Kaiser:2000gs}.
According to Eq.~(\ref{Dmupsi}), the expression $D_\mu\psi D^\mu\psi$ implicitly involves two flavor traces
(see footnote 7 of Ref.~\cite{Kaiser:2000gs}), with the result that the corresponding term is ${\cal O}(N_c^0)$.

The SU(3) Lagrangian of ${\cal O}(p^6)$ was discussed in
Refs.~\cite{Fearing:1994ga,Bijnens:1999sh,Ebertshauser:2001nj,Bijnens:2001bb}, and the generalization
to the U(3) case has recently been obtained in Ref.~\cite{Jiang:2014via}.
For the present purposes, at NNLO, the relevant pieces of $\mathcal{L}^{(2)}$ can be split into three different contributions
of ${\cal O} (N_c^{-1}p^2)$, ${\cal O}(p^4)$, and ${\cal O}(N_c p^6)$, respectively:
\begin{align}
	\label{L2a}
	{\cal L}^{(2, N_c^{-1}p^2)}&=-\frac{F^2}{4}v^{(2)}_2(\psi+\theta)^2\langle \chi U^\dagger+U\chi^\dagger \rangle,\\
	\label{L2b}
	{\cal L}^{(2, p^4)} &=
	L_4 \langle D_\mu U  D^\mu U^\dagger \rangle \langle \chi U^\dagger + U \chi^\dagger\rangle
	+L_6 \langle \chi U^\dagger+U\chi^\dagger\rangle^2
	+L_7 \langle  \chi U^\dagger - U \chi^\dagger\rangle^2 \nonumber\\
	&\quad + i L_{18} D_\mu \psi \langle  \chi D^\mu U^\dagger - D^\mu U \chi^\dagger \rangle
	+ i L_{25} (\psi+\theta) \langle  \chi U^\dagger \chi U^\dagger- U \chi^\dagger U \chi^\dagger\rangle
	+\ldots,\\
	\label{L2c}
	{\cal L}^{(2,N_cp^6)} &=
	C_{12} \langle \chi_{+}h_{\mu\nu}h^{\mu\nu} \rangle
	+ C_{14} \langle u_{\mu}u^{\mu}\chi^2_{+}\rangle+C_{17} \langle \chi_{+}u_{\mu}\chi_{+}u^{\mu}\rangle
	+ C_{19} \langle\chi^3_{+}\rangle\no\\
	&\quad+C_{31} \langle \chi^2_{-}\chi_{+}\rangle+\ldots,
\end{align}
where
\begin{align}
	\chi_\pm&=u^\dagger\chi u^\dagger\pm u\chi^\dagger u,\nonumber\\
	u&=\sqrt{U},\nonumber\\
	u_{\mu}&=i\left[u^\dagger(\partial_\mu-ir_\mu)u-u(\partial_\mu-il_\mu)u^\dagger\right]=iu^\dagger D_\mu U u^\dagger,\nonumber\\
	h_{\mu\nu}&=\nabla_\mu u_\nu+\nabla_\nu u_\mu,\nonumber\\
	\nabla_{\mu}X&=\partial_\mu X+[\Gamma_\mu , X],\nonumber\\
	\Gamma_{\mu}&=\frac{1}{2}\left[u^\dagger(\partial_\mu-ir_\mu)u+u(\partial_\mu-il_\mu)u^\dagger\right].
	\label{eq:bblocks}
\end{align}
The coupling $v^{(2)}_2$ of Eq.~(\ref{L2a}) scales like ${\cal O}(N_c^{-2})$ and originates from
the expansion of the potentials of Refs.~\cite{Leutwyler:1996sa,Kaiser:2000gs} up to and including
terms of order $(\psi+\theta)^2$.
The first three terms of Eq.~(\ref{L2b}) stem from the standard SU(3) ChPT Lagrangian of ${\cal O}(p^4)$
with two traces and are $1/N_c$ suppressed compared to the single-trace terms in Eq.~(\ref{L1}).
Finally, the $C_i$ terms of Eq.~(\ref{L2c}) are obtained from single-trace terms of the SU(3) Lagrangian of
${\cal O}(p^6)$ \cite{Bijnens:1999sh}.

\subsection{Odd-intrinsic-parity Lagrangians}

The WZW term accounts for the anomaly. Beyond the WZW action, the terms of the odd-intrinsic-parity sector are ordinary local Lagrangians expressible as closed expressions in $U$. However, also in this case, the U(3) unnatural-parity Lagrangian contains additional terms in comparison with its SU(3) counterpart. 
At $\mathcal{O}(p^4)$, there exist six independent invariants which obey charge conjugation invariance, and the effective Lagrangian at $\mathcal{O}(p^4)$ reads \cite{Kaiser:2000gs}
\begin{align}
{\cal L}^{(p^4)}_\epsilon&={\cal L}_{\mbox{\tiny WZW}}+
\tilde{V}_1\,  i \br \tilde{R}^{\mu \nu } D_\mu U D_\nu
U^\dega +  \tilde{L}^{\mu \nu } D_\mu U^\dega D_\nu U \ke
+\tilde{V}_2\, \br \tilde{R}^{\mu \nu } U L_{\mu \nu }U^\dega \ke\no\\
&\quad
+ \tilde{V}_3\, \br \tilde{R}^{\mu \nu } R_{\mu \nu } +\tilde{L}^{\mu \nu }
L_{\mu \nu }   \ke  
+  \tilde{V}_4\,   i D_\mu \theta\,  \br
  \tilde{R}^{\mu \nu } D_\nu U U^\dega -\tilde{L}^{\mu \nu } U^\dega  D_\nu
  U\ke  
\no\\&\quad +\tilde{V}_5
\,( \br \tilde{R}^{\mu \nu } \ke 
\br  R_{\mu \nu } \ke+  \br\tilde{L}^{\mu \nu }
\ke \br L_{\mu \nu } \ke) 
+\tilde{V}_6\, \br \tilde{R}^{\mu \nu } \ke \br  L_{\mu \nu } \ke,
\label{InvariantsWZW}
\end{align}
where
\begin{align}
R_{\mu\nu}&=\partial_\mu r_\nu-\partial_\nu r_\mu-i[r_\mu,r_\nu],\no\\
L_{\mu\nu}&=\partial_\mu l_\nu-\partial_\nu l_\mu-i[l_\mu,l_\nu],\no\\
\tilde{F}^{\mu\nu}&=\frac{1}{2}\epsilon^{\mu\nu\rho\sigma}F_{\rho\sigma},\no\\
\epsilon_{0123}&=1.
\end{align}
Due to parity, all potentials are odd functions of $(\psi+\theta)$, except for $\tilde{V}_4$ which is even. 

In the combined L$N_c$ and chiral expansions, the WZW term starts contributing at $\mathcal{O}(N_cp^4)=\mathcal{O}(\delta)$. 
Our aim is the calculation of the two-photon decays at the one-loop level, which corresponds to a NNLO calculation in the $\delta$ counting. Therefore, we need the odd-intrinsic-parity Lagrangians at NLO and NNLO. Up to and including NNLO, the effective odd-intrinsic-parity Lagrangian is denoted by
\begin{align}
\mathcal{L}_\epsilon=\mathcal{L}^{(1)}_{\text{WZW}}+\mathcal{L}^{(2)}_\epsilon+\mathcal{L}^{(3)}_\epsilon,
\end{align}
where the superscripts $(i)$ refer to the order in $\delta$.
The NLO Lagrangian $\mathcal{L}^{(2)}_\epsilon$ receives contributions from $\mathcal{O}(p^4)$ and $\mathcal{O}(N_cp^6)$. From Eq.~(\ref{InvariantsWZW}) one can extract \cite{Kaiser:2000gs}
\begin{align}
{\cal L}^{(2,p^4)}_\epsilon&=
\tilde{L}_1\,i\hspace{0.1 em}(\psi+\theta) \br \tilde{R}^{\mu \nu } D_\mu U D_\nu
U^\dega+ \tilde{L}^{\mu\nu} D_\mu U^\dega D_\nu U \ke \no\\
&\quad + \tilde{L}_2(\psi+\theta)\br \tilde{R}^{\mu\nu} U L_{\mu \nu }U^\dega \ke+ \tilde{L}_3(\psi+\theta)\br \tilde{R}^{\mu \nu } R_{\mu \nu  }
+ \tilde{L}^{\mu \nu } L_{\mu \nu  } \ke \no\\ 
&\quad
+  \tilde{L}_4\,   i D_\mu \theta\,  \br
 \tilde{R}^{\mu \nu} D_\nu U U^\dega - \tilde{L}^{\mu \nu }U^\dega  D_\nu U\ke.
\end{align}
The odd-intrinsic-parity Lagrangian at $\mathcal{O}(p^6)$ has been constructed in SU(3) ChPT in Refs.~\cite{Ebertshauser:2001nj,Bijnens:2001bb}. Reference \cite{Jiang:2014via} provides the full $\mathcal{O}(p^6)$ Lagrangian in U(3) ChPT. The $\mathcal{O}(N_cp^6)$ contributions are those terms of the $\mathcal{O}(p^6)$ Lagrangian which have only one flavor trace and do not contain the fields $(\psi+\theta)$ or $D_\mu \theta$. The NNLO Lagrangian $\mathcal{L}^{(3)}_\epsilon$ consists of terms of $\mathcal{O}(N_cp^8)$, $\mathcal{O}(p^6)$, and $\mathcal{O}(N^{-1}_c p^4)$. The $\mathcal{O}(p^8)$ Lagrangian has not been constructed so far. Order $p^6$ terms stem from the $\mathcal{O}(p^6)$ Lagrangian containing two flavor traces, the field $D_\mu \theta$, or are generated when expanding the potentials of the odd-parity terms of the $\mathcal{O}(p^6)$ Lagrangian up to and including linear order in $(\psi+\theta)$. Terms of $\mathcal{O}(N^{-1}_c p^4)$ could arise from the expansion of the potentials in ${\cal L}^{(p^4)}_\epsilon$, but they do not contribute to the two-photon decays. Since especially the $\mathcal{O}(p^6)$ Lagrangian contains a lot of terms, we display only the terms needed for the calculations in this work. The relevant terms of the $\mathcal{O}(N_cp^6)$ and $\mathcal{O}(p^6)$ Lagrangians are shown in Table \ref{tab:TermsNcp6}. Since there is, at present, no satisfactory unified nomenclature for the coupling constants, for easier reference we choose the names according to the respective references from which the Lagrangians were taken.

The operators with the LECs $L^{6,\epsilon}_i$, which appear in SU(3) ChPT as well, are taken from Ref.~\cite{Ebertshauser:2001nj}. They are given in terms of the building blocks
\begin{align}
(A)_{\pm}&=u^\dagger A u^\dagger\pm u A^\dagger u,\no\\
G^{\mu\nu}&=R^{\mu\nu} U+U L^{\mu\nu},\no\\
H^{\mu\nu}&=R^{\mu\nu} U- U L^{\mu\nu},\no\\
(D_\mu D_\nu U)^s_-&=\frac{1}{2}(\{D_\mu,D_\nu\}U)_-\no\\
&=(D_\mu D_\nu U)_-+\frac{i}{2}(H_{\mu\nu})_+,
\end{align}
where $A$ refers to operators transforming under the chiral group $G$ as $A\stackrel{G}{\rightarrow}V_R A V^\dagger_L$.
The other terms, genuinely related to the U(3) sector, are taken from Ref.~\cite{Jiang:2014via}. Here, the corresponding building blocks are the same as in Eq.~(\ref{eq:bblocks}) with the additional structures
\begin{align}
f^{\mu\nu}_\pm=uL^{\mu\nu} u^\dagger\pm u^\dagger R^{\mu\nu} u.
\end{align}

\begin{table}[htbp]
	\hspace{-0.75cm}
		\begin{tabular}{c c c c c}\hline\hline%\toprule\toprule
		Lagrangian& LEC& Operator &SU(3)\\
		\hline%\midrule
		$\mathcal{L}^{(2,N_c p^6)}_\epsilon$& $L^{6,\epsilon}_3$ & $ i\ \langle (\chi)_+ \{ (G_{\mu\nu})_+ (H_{\alpha\beta})_+ -  \mbox{rev} \} \rangle\epsilon^{\mu\nu\alpha\beta}$ & x\\
		& $L^{6,\epsilon}_8$&$ i\ \langle (\chi)_- (G_{\mu\nu})_+ (G_{\alpha\beta})_+ \rangle\epsilon^{\mu\nu\alpha\beta}$ & x\\
		& $L^{6,\epsilon}_{19} $&$i\ \langle (D^\lambda G_{\lambda\mu})_+ \{ (G_{\nu\alpha})_+ (D_\beta U)_- + \mbox{rev} \}\rangle\epsilon^{\mu\nu\alpha\beta}$  & x\\ \hline%\midrule
$\mathcal{L}^{(3,p^6)}_\epsilon$& $L^{6,\epsilon}_9$&$ i\ \langle (\chi)_- \rangle\langle (G_{\mu\nu})_+ (G_{\alpha\beta})_+ \rangle\epsilon^{\mu\nu\alpha\beta}$ & x\\
		& $ L_{237}$ &$\epsilon^{\mu\nu\lambda\rho}\la f_{+\mu\nu}\ra\la{f_{+\lambda}}^{\sigma}h_{\rho\sigma}\ra$ & -\\
		& $L_{238}$ &$ \epsilon^{\mu\nu\lambda\rho}\la f_{+\mu\nu}\ra\la\nabla^{\sigma}f_{+\lambda\sigma}u_{\rho}\ra$ & -\\
		& $L_{239} $&$\epsilon^{\mu\nu\lambda\rho}\la f_{+\mu\nu}\nabla^{\sigma}f_{+\lambda\sigma}\ra\la u_{\rho}\ra$ & -\\
		& $L_{258}$&$ i \epsilon^{\mu\nu\lambda\rho}\la f_{+\mu\nu}\ra\la f_{+\lambda\rho}\chim\ra$ & - \\
		&$\Lambda_{442} $&$\epsilon^{\mu\nu\lambda\rho}(\psi+\theta)\la f_{+\mu\nu}f_{+\lambda\rho}\chip\ra$& -\\ 
		\hline\hline%\bottomrule\bottomrule
		\end{tabular}
			\caption{Relevant terms of $\mathcal{L}^{(2,N_c p^6)}_\epsilon$ and $\mathcal{L}^{(3,p^6)}_\epsilon$.}
	\label{tab:TermsNcp6}
\end{table}

\subsection{Power counting}

   In the following, we provide the power-counting rules for a given Feynman diagram, which has been evaluated by using the
interaction vertices derived from the effective Lagrangians of Eq.~(\ref{Leff}).
   Using the $\delta$ counting introduced in Eq.~(\ref{powerexp}), we assign
to any such diagram an order $D$ which is obtained from the following ingredients:
   Meson propagators for both octet and singlet fields count as $\mathcal{O}(\delta^{-1})$.
   Since meson fields are always divided by $F=\mathcal{O}(\sqrt{N_c})=\mathcal{O}(\delta^{-\frac{1}{2}})$, a
vertex with $k$ meson fields derived from $\mathcal{L}^{(i)}$ is $\mathcal{O}(\delta^{i+k/2})$.
   The integration of a loop counts as $\delta^2$.
   The order $D$ is obtained by adding up the contributions of the individual building blocks.
   The power-counting rules are summarized in Table~\ref{TabCounting}.

\begin{table}[htbp]
\tabcolsep2em
\renewcommand{\arraystretch}{1.2}
\begin{tabular}{l c c c}
\hline\hline
Quantity &$N_c$ & $p$ & $\delta$ \\ \hline
Momenta/Derivatives $p$/$\partial_{\mu}$ & $1$ & $p$ & $\delta^\frac{1}{2}$ \\
$1/N_c$ & $N_c^{-1}$ & $1$ & $\delta$ \\
Quark masses $m$ & $1$ & $p^2$ & $\delta$ \\
Dynamical fields $\phi_a$ ($a=1,\ldots, 8$)& $\sqrt{N_c}$ & $1$ & $\delta^{-\frac{1}{2}}$\\
Dynamical field $\psi$ & $1$ & $1$ & $1$\\
External field $\theta$ & $1$ & $1$ & $1$\\
External currents $v_{\mu}$ and $a_{\mu}$ & $1$ & $p$ & $\delta^\frac{1}{2}$ \\
External fields $s$ and $p$ & $1$ & $p^2$ & $\delta$\\
Pion-decay constant $F$ (chiral limit) & $\sqrt{N_c}$ & $1$ & $\delta^{-\frac{1}{2}}$ \\
Topological susceptibility $\tau$  & 1 & 1 &  1\\
$M^2_{\eta'}$ (chiral limit)& $N_c^{-1}$& $1$ & $\delta$\\
Octet-meson propagator & $1$& $p^{-2}$ & $\delta^{-1}$\\
Singlet-$\eta_1$ propagator (chiral limit) & a) & a) & $\delta^{-1}$\\
Loop integration & $1$& $p^4$ & $\delta^2$\\
$k$-meson vertex from ${\cal L}^{(i)}$ & b) & b) & $\delta^{i+k/2}$\\
\hline\hline
\end{tabular}
\caption{Power-counting rules in L$N_c$ChPT.
a) The inverse of the singlet $\eta_1$ propagator is of order $1/N_c$ and $p^2$.
b) The assignment $i$ in ${\cal L}^{(i)}$ receives contributions from both
$1/N_c$ and $p^2$.
   Recall that powers $(\psi+\theta)^n$ come with expansion coefficients of ${\cal O}(N_c^{-n})$ even
though we count $(\psi+\theta)$ as ${\cal O}(1)$.
}
\label{TabCounting}
\end{table}

\section{Calculation of the invariant amplitude}
\label{sec:TPDCalcA}

\begin{figure}
	\centering
		\includegraphics[width=0.98\textwidth]{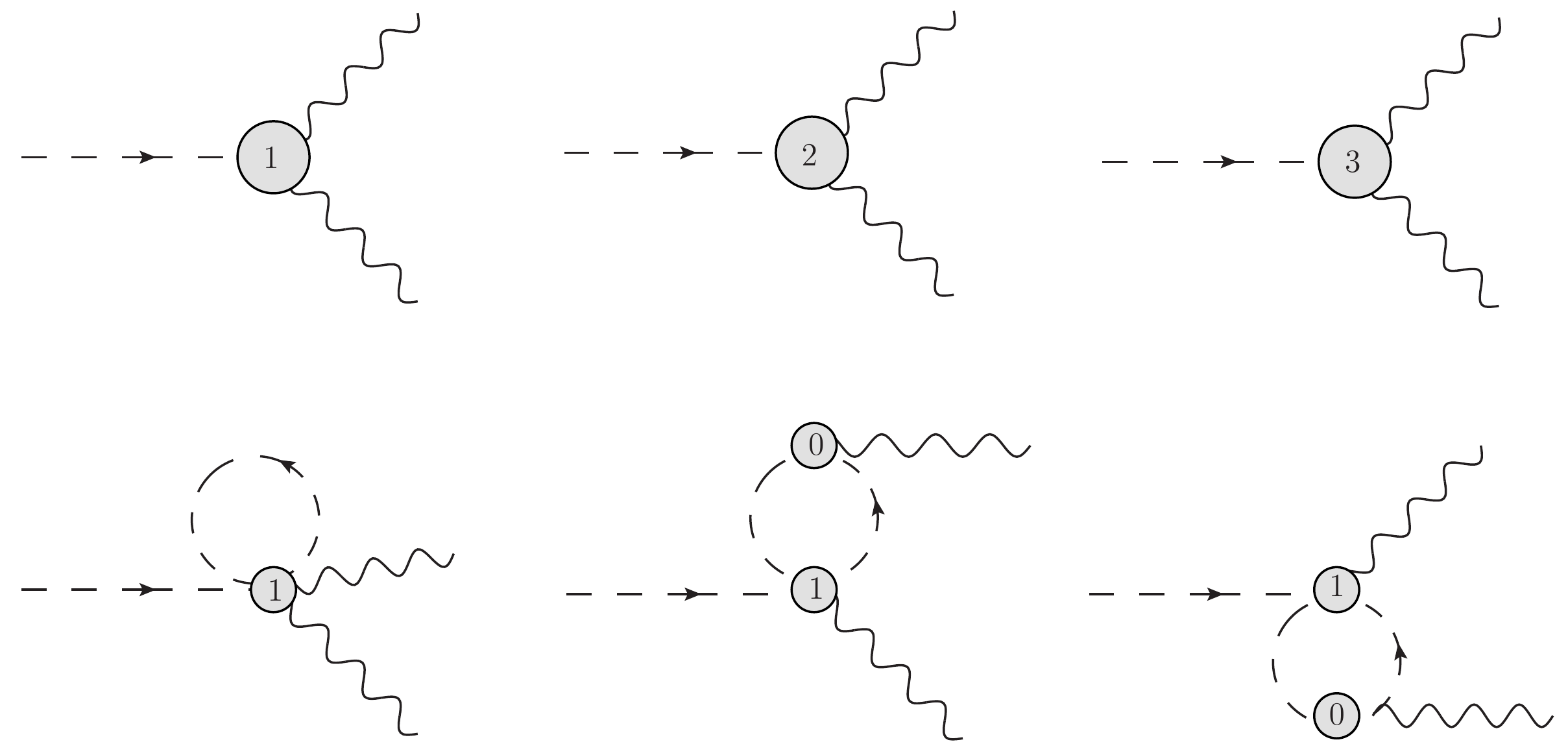}
	\caption{Feynman diagrams for $P\to\gamma^*\gamma^*$ up to and including NNLO. Dashed lines refer to pseudoscalar mesons and wiggly lines to photons. The numbers $k$ in the interaction blobs refer to vertices derived from the corresponding Lagrangians $\mathcal{L}^{(k)}$.}
	\label{fig:phiggallp}
\end{figure}

The invariant amplitude of the two-photon decay of a pseudoscalar meson $P$ can be parameterized by
\begin{align}
\mathcal{M}=-iF_{P\gamma^*\gamma^*}(q^2_1,q^2_2)\epsilon_{\mu\nu\alpha\beta}\epsilon^\mu_1\epsilon^\nu_2 q^\alpha_1 q^\beta_2,
\label{matrixeTPD}
\end{align}
where $q^\mu_1$, $q^\mu_2$ denote the photon momenta, $\epsilon^\mu_1$, $\epsilon^\mu_2$ 
the polarization vectors of the photons, and $F_{P\gamma^*\gamma^*}(q^2_1,q^2_2)$ is the so-called transition form factor (TFF). In order to determine the invariant amplitude up to and including NNLO, we need to calculate the Feynman diagrams shown in Fig.~\ref{fig:phiggallp}.
The vertices are derived from the Lagrangians given in Sec.~\ref{sec:Lagrangians}. The coupling of the electromagnetic field to the mesons is described by introducing an external field which couples to the electromagnetic current operator
\begin{align}
J^\mu=\bar{q}Q\gamma^\mu q,
\end{align}
where $Q$ is the quark-charge matrix. For $N_c=3$, the quark-charge matrix is given by
\begin{align}
Q(3)=\text{diag}\left(\frac{2}{3},-\frac{1}{3},-\frac{1}{3}\right).
\end{align}
However, as B\"ar and Wiese pointed out \cite{Bar:2001qk}, in order for the Standard Model to be consistent for arbitrary $N_c$, the ordinary quark-charge matrix should be replaced by
\begin{align}
Q(N_c)=\frac{1}{2}\text{diag}\left(\frac{1}{N_c}+1,\frac{1}{N_c}-1,\frac{1}{N_c}-1\right).
\end{align}
The matrix element $\mathcal{M}$ is calculated using both versions, $Q(3)$ and $Q(N_c)$. In the $Q(N_c)$ case, we first perform the $\delta$ expansion up to and including NNLO and then set $N_c=3$. The Feynman diagrams are evaluated using the \textit{Mathematica} package FEYNCALC \cite{Mertig:1990an}.

For the decays of $\eta$ and $\eta'$, we take into account the $\eta$-$\eta'$ mixing at NNLO\footnote{No mixing with $\pi^0$.}. A detailed derivation of the $\eta$-$\eta'$ mixing at NNLO can be found in Ref.~\cite{Bickert:2016fgy}. First, we calculate the coupling of two photons to the octet and singlet fields $\phi_b$, collected in the doublet $\eta_A\equiv\left(\eta_8,\eta_1\right)^T$, at the one-loop level up to and including NNLO in the $\delta$ counting. The result, which should be interpreted as a Feynman rule, is represented by the ``matrix elements'' $\mathcal{F}_{b}=\left\langle \gamma^*\gamma^*|b\right\rangle$. In a next step, we transform the bare fields $\eta_A$ to the physical states using the transformation $T$ in Eq.~(51) in Ref.~\cite{Bickert:2016fgy}:
\begin{align}
\begin{pmatrix} \eta_8 \\
	\eta_1
\end{pmatrix}=
\begin{pmatrix} T_{8\eta} & T_{8\eta'}\\
	T_{1\eta} & T_{1\eta'}
\end{pmatrix}
\begin{pmatrix} \eta \\
	\eta'
\end{pmatrix}.
\end{align}
The resulting (``physical'') matrix elements are then given by
\begin{align}
	\begin{pmatrix} F_{\eta\gamma^\ast\gamma^\ast} \\
		F_{\eta'\gamma^\ast\gamma^\ast}
	\end{pmatrix}=
	\begin{pmatrix} T_{8\eta} & T_{1\eta}\\
		T_{8\eta'} & T_{1\eta'}
	\end{pmatrix}
	\begin{pmatrix} \mathcal{F}_8 \\
		\mathcal{F}_1
	\end{pmatrix}.
\end{align}
For the calculation of the loop diagrams, we employ the LO mixing. 

Without the $1/N_c$ expansion of $Q$, the results for the form factors of $\pi^0$, $\eta$, $\eta'$ at LO and NLO read
\begin{align}
&F_{\pi^0\gamma^*\gamma^*}^\text{LO}=\frac{1}{4 \pi ^2 F_{\pi }},\\
&F_{\pi^0\gamma^*\gamma^*}^\text{NLO}=\frac{1}{4 \pi ^2 F_{\pi }}\left[1-\frac{1024}{3} \pi ^2 M_{\pi }^2 L_8^{6, \epsilon }-\frac{512}{3} \pi ^2 L_{19}^{6, \epsilon }\left(q^2_1+q^2_2\right)\right],\\
&F_{\eta\gamma^*\gamma^*}^\text{LO}=\frac{1}{4 \sqrt{3} \pi ^2 F_{\pi }}\left[\cos (\theta^{[0]})-2 \sqrt{2} \sin (\theta^{[0]})\right],\\
&F_{\eta\gamma^*\gamma^*}^\text{NLO}\no\\
&=\frac{1}{4 \sqrt{3} \pi ^2 F_{\pi }}\left\{\cos (\theta^{[1]})-2 \sqrt{2} \sin (\theta^{[1]})\right.\no\\
&\quad+\frac{8 \left(M_K^2-M_{\pi }^2\right) \left[\sqrt{2} \sin (\theta^{[1]})+2 \cos (\theta^{[1]})\right]}{3 F_{\pi }^2}L_5\no\\
&\quad+\frac{1024}{9} \pi ^2 \left[2 \sqrt{2} \left(M_K^2+2 M_{\pi }^2\right)
   \sin (\theta^{[1]})+\left(4 M_K^2-7 M_{\pi }^2\right) \cos (\theta^{[1]})\right]L^{6, \epsilon}_8\no\\
&\quad+\sqrt{2} \sin (\theta^{[1]}) \lambda_1\no\\
&\quad\left.-\frac{512}{3} \pi ^2 \left[\cos (\theta^{[1]}) -2 \sqrt{2} \sin (\theta^{[1]}) \right]L_{19}^{6, \epsilon }\left(q^2_1+q^2_2\right)\right\},\\
&F_{\eta'\gamma^*\gamma^*}^\text{LO}=\frac{1}{4 \sqrt{3} \pi ^2 F_{\pi }}\left[\sin (\theta^{[0]})+2 \sqrt{2} \cos (\theta^{[0]})\right],\\
&F_{\eta'\gamma^*\gamma^*}^\text{NLO}\no\\
&=\frac{1}{4 \sqrt{3} \pi ^2 F_{\pi }}\left\{\sin (\theta^{[1]})+2 \sqrt{2} \cos (\theta^{[1]})\right.\no\\
&\quad+\frac{8 \left(M_{\pi }^2-M_K^2\right) \left[\sqrt{2} \cos (\theta^{[1]})-2 \sin (\theta^{[1]})\right]}{3 F_{\pi }^2}L_5\no\\
&\quad-\frac{1024}{9} \pi ^2 \left[\left(7 M_{\pi }^2-4 M_K^2\right) \sin
   (\theta^{[1]})+2 \sqrt{2} \left(M_K^2+2 M_{\pi }^2\right) \cos (\theta^{[1]})\right]L^{6, \epsilon}_8\no\\
	&\quad-\sqrt{2} \cos (\theta^{[1]}) \lambda_1\no\\
	&\quad\left.-\frac{512}{3} \pi ^2 \left[\sin (\theta^{[1]}) +2 \sqrt{2} \cos(\theta^{[1]}) \right]L_{19}^{6, \epsilon }\left(q^2_1+q^2_2\right)\right\},
\end{align}
where $\theta^{[i]}$ is the corresponding mixing angle at LO (NLO) obtained from Eq.~(49) in Ref.~\cite{Bickert:2016fgy}. The parameter $\lambda_1$ is a QCD-scale-invariant combination of parameters violating the Okubo-Zweig-Iizuka (OZI) rule, given by \cite{Kaiser:2000gs}
\begin{align}
\lambda_1=\Lambda_1-2K_1=\Lambda_1+16\pi^2(\tilde{L}_2+2\tilde{L}_3).
\end{align}
Including the $1/N_c$ expansion of $Q$, the results at LO and NLO now take the form
\begin{align}
&F_{\pi^0\gamma^*\gamma^*}^\text{LO}=0,\\
&F_{\pi^0\gamma^*\gamma^*}^\text{NLO}=\frac{1}{4 \pi ^2 F_{\pi }},\\
&F_{\eta\gamma^*\gamma^*}^\text{LO}=-\frac{3 \sqrt{\frac{3}{2}}}{8 \pi ^2 F_{\pi }} \sin (\theta^{[0]}),\\
&F_{\eta\gamma^*\gamma^*}^\text{NLO}\no\\
&=\frac{3 \sqrt{\frac{3}{2}}}{8 \pi ^2 F_{\pi }}\left\{- \sin (\theta^{[1]})\right.\no\\
&\quad+\frac{8 \left(M_K^2-M_{\pi }^2\right) \left[\sin (\theta^{[1]})+\sqrt{2} \cos (\theta^{[1]})\right]}{3
   F_{\pi }^2}L_5\no\\
&\quad+\frac{1024}{9} \pi ^2 \left[\left(2 M_K^2+M_{\pi }^2\right) \sin (\theta^{[1]})+2 \sqrt{2}
   \left(M_K^2-M_{\pi }^2\right) \cos (\theta^{[1]})\right]L^{6, \epsilon}_8\no\\
&\quad\left.+\frac{\sin (\theta^{[1]})}{2}\lambda_1+384
   \sqrt{2} \pi ^2 \sin (\theta^{[1]}) L_{19}^{6, \epsilon }\left(q^2_1+q^2_2\right)\right\},\\
&F_{\eta'\gamma^*\gamma^*}^\text{LO}=\frac{3 \sqrt{\frac{3}{2}} }{8 \pi ^2 F_{\pi }}\cos (\theta^{[0]}),\\
&F_{\eta'\gamma^*\gamma^*}^\text{NLO}\no\\
&=\frac{3 \sqrt{\frac{3}{2}}}{8 \pi ^2 F_{\pi }}\left\{ \cos (\theta^{[1]})\right.\no\\
&\quad+\frac{8 \left(M_K^2-M_{\pi }^2\right) \left[\sqrt{2} \sin (\theta^{[1]})-\cos (\theta^{[1]})\right]}{3
   F_{\pi }^2}L_5\no\\
&\quad+\frac{1024}{9} \pi ^2 \left[2 \sqrt{2} \left(M_K^2-M_{\pi }^2\right) \sin (\theta^{[1]}) -\left(2 M_K^2+M_{\pi }^2\right) \cos (\theta^{[1]})\right]L^{6, \epsilon}_8\no\\
&\quad\left.-\frac{\cos (\theta^{[1]})}{2}\lambda_1-384
   \sqrt{2} \pi ^2 \cos (\theta^{[1]}) L_{19}^{6, \epsilon }\left(q^2_1+q^2_2\right)\right\}.
\end{align}
At NNLO, the expressions for the form factors are quite long. Therefore, we do not display all terms explicitly. The loop contributions corresponding to the loop diagrams shown in Fig.~\ref{fig:phiggallp} are provided in Appendix \ref{app:expAnomD}.  
The expressions for the full NNLO form factors, with tree-level contributions, are available as \textit{Mathematica} notebooks.

\subsection{Observables}

The decay amplitude for real photons is recovered by setting $q^2_1=q^2_2=0$ in Eq. (\ref{matrixeTPD}). The decay width is then given by \cite{Hacker:2008}
\begin{align}
\Gamma=\frac{1}{2!2M_P(2\pi)^2}\frac{\pi\sqrt{\lambda[M^2_P,0,0]}}{2M^2_P}\int \text{d}\Omega\sum_{\lambda_1,\lambda_2}|\mathcal{M}|^2,
\end{align}
where $\lambda(x,y,z)=x^2+y^2+z^2-2xy-2yz-2xz$ is the K\"aell\'en function, $\lambda_1,\lambda_2$ denote the polarizations of the photons, and $\text{d}\Omega$ is the solid angle of one of the photons.
Using $\sum_{\lambda}\epsilon^*_{(\lambda)\mu}\epsilon_{(\lambda)\mu'}=-g_{\mu\mu'}$, one obtains
\begin{align}
\Gamma(P\to\gamma\gamma)=\frac{M^3_P}{64\pi}|F_{P\gamma\gamma}|^2.
\end{align}
The single-virtual TFF $F_{P\gamma^*\gamma}(q^2):=F_{P\gamma^*\gamma^*}(q^2,0)$ can be measured in single Dalitz decays $P\to\gamma l^+l^-$. 
The $slope$ of the TFF is defined as
\begin{align}
slope\ :=\frac{1}{F_{P\gamma\gamma}}\frac{d}{dq^2}\left.F_{P\gamma^*\gamma}(q^2)\right|_{q^2=0}.
\label{eq:slope}
\end{align}
One can also define the dimensionless quantity $b_P=M^2_P\times slope$.
The curvature is given by
\begin{align}
curv\ :=\frac{1}{2}\frac{1}{F_{P\gamma\gamma}}\frac{d^2}{d(q^2)^2}\left.F_{P\gamma^*\gamma}(q^2)\right|_{q^2=0},
\label{eq:curv}
\end{align}
and the corresponding dimensionless quantity reads $c_P=M^4_P\times curv$.

Experimental extractions of the slope parameter are often performed using a vector-meson-dominance model (VMD) \cite{Sakurai:69} to fit the data. Introducing $G_{P\gamma^*\gamma}(Q^2)=F_{P\gamma^*\gamma}(q^2)$ with $Q^2=-q^2$, in this case, the TFF is given by a normalized single-pole term with an associated mass $\Lambda_P$ \cite{Escribano:2013kba}:
\begin{align}
G_{P\gamma^*\gamma}(Q^2)=\frac{F_{P\gamma\gamma}(0)}{1+Q^2/\Lambda^2_P}.
\end{align}
Expanding this expression in $Q^2$ leads to
\begin{align}
G_{P\gamma^*\gamma}(Q^2)=F_{P\gamma\gamma}(0)\left(1-\frac{Q^2}{\Lambda^2_P}+\frac{(Q^2)^2}{\Lambda^4_P}+\dots\right).
\end{align}
Now, we can read off the slope and curvature VMD predictions, which are given by
\begin{align}
slope&=\frac{1}{\Lambda^2_P},\\
curv&=\frac{1}{\Lambda^4_P}.
\label{eq:slopecurvVMD}
\end{align}

\section{Numerical analysis}

We perform the numerical analysis of our results successively at LO, NLO, and NNLO. In the following, we distinguish between two cases: (a) using the normal quark-charge matrix for $N_c=3$ and (b) taking the $1/N_c$ expansion of $Q$ into account, denoted by Qexp. Performing the $1/N_c$ expansion of $Q$ shifts some of the LECs to higher orders. The LECs $L^{6,\epsilon}_8$ and  $L^{6,\epsilon}_{19}$, stemming from the NLO Lagrangian in Table \ref{tab:TermsNcp6} in Sec.~\ref{sec:Lagrangians}, appear only at NNLO in the expression for the two-photon decay of the $\pi^0$. At LO, no unknown LECs show up and we can calculate the desired quantities directly.

\subsection{NLO}
\label{sec:tpdNLO}

At NLO, we have to determine five LECs. From the even-intrinsic-parity sector, $L_5$ and the NLO mixing angle $\theta^{[1]}$ contribute. Here, we employ the values for $L_5$ and $\theta^{[1]}$ determined in the NLO analysis of the $\eta$-$\eta'$ mixing in Tables II and IV in Ref.~\cite{Bickert:2016fgy} labeled NLO I, namely, $ L_5=(1.86\pm 0.06)\times 10^{-3}$ and $\theta^{[1]}=(-11.6\pm 0.6)\deg$. From the odd-intrinsic-parity sector we have to fix $L^{6,\epsilon}_8$, $L^{6,\epsilon}_{19}$, and $\lambda_1=\Lambda_1-2K_1$. 

\subsubsection{Determinations of the parameters}
First, we consider the $Q(3)$ case. Since the decay width of the $\pi^0$ to two photons depends only on $L^{6,\epsilon}_8$, we start by fixing $L^{6,\epsilon}_8$ to the experimental value of $\Gamma_{\pi^0\to\gamma\gamma}$. We then fit $\lambda_1$ simultaneously to the experimental results for $\Gamma_{\eta\to\gamma\gamma}$ and $\Gamma_{\eta'\to\gamma\gamma}$. The experimental values for the decay widths are taken from Ref.~\cite{Tanabashi:2018oca} and are displayed in Table \ref{tab:WidthsNLO}. Finally, the parameter $L^{6,\epsilon}_{19}$ is determined through a simultaneous fit to the experimental values of the $\pi^0$, $\eta$, and $\eta'$ slopes, given in Table \ref{tab:WidthsNLO}. For the fits we employ the \textit{Mathematica} routine NonlinearModelFit. The errors of the fit parameters and of the results are the ones obtained from the fit routine. The different steps are performed successively and we do not take the errors of LECs determined in a previous step into account. We also do not consider the errors due to neglecting higher-order terms. In principle, a systematic error of at least 10\%, corresponding to $\delta^2=1/9$, should be added to all quantities determined up to NLO. 
The results for the LECs are given in Table \ref{tab:tpdNLO} and the results for the decay widths and slopes in Table \ref{tab:WidthsNLO}, labeled NLO 1.

Next, we examine the case where we fit $L^{6,\epsilon}_8$ and $\lambda_1$ simultaneously to all three decay widths  $\Gamma_{\pi^0\to\gamma\gamma}$, $\Gamma_{\eta\to\gamma\gamma}$, and $\Gamma_{\eta'\to\gamma\gamma}$. The constant $L^{6,\epsilon}_{19}$ is then again fixed to the slopes of $\pi^0$, $\eta$, and $\eta'$. The results are shown in Table \ref{tab:tpdNLO}, labeled NLO 2. 
To consistently take the errors of $L^{6,\epsilon}_8$, $L^{6,\epsilon}_{19}$, and $\lambda_1$ into account, we consider another scenario where we determine these three LECs through a simultaneous fit to the decay widths of $\pi^0$, $\eta$, $\eta'$ and the slope parameters of $\pi^0$, $\eta$, $\eta'$. The results are given in Table \ref{tab:tpdNLO}, labeled NLO 3. 

In the $Q(N_c)$ case, the $\pi^0$ form factor at NLO is independent of $L^{6,\epsilon}_8$ and  $L^{6,\epsilon}_{19}$, which are shifted to the NNLO expression. The width $\Gamma_{\pi^0\to\gamma\gamma}$ takes the LO value of the $Q(3)$ case, and the slope is equal to zero at NLO. Therefore, we determine $L^{6,\epsilon}_8$, $L^{6,\epsilon}_{19}$, and $\lambda_1$ via a simultaneous fit to $\Gamma_{\eta\to\gamma\gamma}$, $\Gamma_{\eta'\to\gamma\gamma}$, $b_\eta$, and $b_\eta'$. The results are displayed in Table \ref{tab:tpdNLO}, labeled NLO, Qexp.

The parameters $L_5$ and $\theta^{[1]}$ have been determined in the NLO analysis in Ref.~\cite{Bickert:2016fgy} with a small error. Therefore, we have not taken these errors into account in the analysis of the two-photon decays. However, to obtain an estimate of the effect of the errors, we recalculated the quantities in the NLO 2 scenario varying $L_5$ and $\theta^{[1]}$ within their errors. This led to only small variations in the last digit of the results for the decay widths and the slopes. We conclude that the influence of the errors of $L_5$ and $\theta_1$ is very small, and we omit them in the following.

\begin{table}[tbp]
	\centering
		\begin{tabular}{c r@{$\,\pm\,$}l r@{$\,\pm\,$}l r@{$\,\pm\,$}l}\hline\hline%\toprule\toprule
		 & \multicolumn{2}{c}{$L^{6,\epsilon}_8\ [10^{-3}]$} &\multicolumn{2}{c}{$L^{6,\epsilon}_{19}\ [10^{-3}]$} & \multicolumn{2}{c}{$\lambda_1$} \\ \hline%\midrule
  \text{NLO 1} & $0.16$&$0.17$ & $-1.26$&$0.17$ & $0.04$&$0.12$ \\
 \text{NLO 2} & $0.86$&$0.13$ & $-0.94$&$0.53$ & $2.59$&$0.13$ \\
 \text{NLO 3} & $0.76$&$0.42$ & $-0.92$&$0.44$ & $2.75$&$0.39$ \\
 \text{NLO, Qexp} & $0.23$&$0.42$ & $-0.67$&$1.86$ & $2.16$&$0.82$ \\
\hline\hline%\bottomrule\bottomrule
		\end{tabular}
		\caption{Results for the LECs determined at NLO.}
		\label{tab:tpdNLO}
		\end{table}
		
\begin{table}[tbp]
	\centering
	\small{
		\begin{tabular}{c r@{$\,\pm\,$}l r@{$\,\pm\,$}l r@{$\,\pm\,$}l r@{$\,\pm\,$}l r@{$\,\pm\,$}l r@{$\,\pm\,$}l}\hline\hline%\toprule\toprule
	& \multicolumn{2}{c}{$\Gamma_{\pi^0}$ [eV]}	& \multicolumn{2}{c}{$\Gamma_\eta$ [keV]} &\multicolumn{2}{c}{$\Gamma_{\eta'}$ [keV]} &  \multicolumn{2}{c}{$b_{\pi^0}$}&\multicolumn{2}{c}{$b_{\eta}$} & \multicolumn{2}{c}{$b_{\eta'}$}\\ \hline%\midrule
 \text{LO} & $7.79$&$0.02$ & $0.62$&$0.00$ & $5.03$&$0.01$ & $0$&$0$ & $0$&$0$ & $0$&$0$ \\
 \text{LO, Qexp} & $0$&$0$ & $0.20$&$0.00$ & $8.33$&$0.02$ & $0$&$0$ & $0$&$0$ & $0$&$0$ \\
 \text{NLO 1} & $7.63$&$0.16$ & $0.60$&$0.27$ & $4.20$&$8.36$ & $0.04$&$0.02$ & $0.52$&$0.31$ & $2.36$&$1.39$ \\
 %\text{NLO, Qexp} & $7.79$&$0.02$ & $0.52$&$0.$ & $4.36$&$0.$ & $0.$&$0.$ & $0.12$&$4.25$ & $-1.49$&$52.63$ \\
 \text{NLO 2} & $6.98$&$1.50$ & $0.50$&$0.93$ & $4.34$&$7.30$ & $0.03$&$0.07$ & $0.43$&$1.03$ & $-1.74$&$4.18$ \\
 \text{NLO 3} & $7.08$&$1.20$ & $0.43$&$0.69$ & $4.75$&$5.57$ & $0.03$&$0.05$ & $0.45$&$0.64$ & $-1.63$&$2.61$ \\
 \text{NLO, Qexp} & $7.79$&$0.02$ & $0.52$&$7.29$ & $4.36$&$56.70$ & $0$&$0$ & $0.12$&$4.25$ & $-1.49$&$52.63$ \\
 \text{Data \cite{Tanabashi:2018oca}} & $7.63$&$0.16$ & $0.51$&$0.02$ & $4.36$&$0.14$ & $0.034$&$0.003$ & $0.59$&$0.02$ & $1.49$&$0.16$ \\
	\hline\hline%\bottomrule\bottomrule		
		\end{tabular}}
		\caption{Results for the two-photon decay widths and the slope parameters at NLO.}
		\label{tab:WidthsNLO}
		\end{table}

\subsubsection{Discussion of the results}

In the NLO 1 case, we find quite small values for $L^{6,\epsilon}_8$ and $\lambda_1$. However, if we perform a simultaneous fit to the $\pi^0$, $\eta$, $\eta'$ decay widths (NLO 2 and NLO 3, which yield similar results), the values for $L^{6,\epsilon}_8$ and $\lambda_1$ become larger, with a drastic increase of the $\lambda_1$ value. Phenomenological studies \cite{Leutwyler:1997yr,Feldmann:1998vh,Feldmann:1998sh,Escribano:2015yup} suggest that OZI-rule-violating parameters as, e.g., $\lambda_1$ should be small. For example, Ref.~\cite{Leutwyler:1997yr} determines\footnote{In Ref.~\cite{Leutwyler:1997yr} the coupling $K_1$ is denoted by $\Lambda_3=K_1$.} $\lambda_1=\Lambda_1-2\Lambda_3=0.25$ and Ref.~\cite{Escribano:2015yup} finds $\Lambda_1=0.21(5)$, $\Lambda_3=0.05(3)$, yielding $\lambda_1=\Lambda_1-2\Lambda_3=0.11(8)$. These results are in agreement with the NLO 1 case, whereas the scenarios NLO 2 and NLO 3 indicate very large OZI-rule-violating corrections. 
The values for $L^{6,\epsilon}_{19}$ do not exhibit large variations in the different scenarios. They can be compared to a VMD prediction yielding $L^{6,\epsilon}_{19}=-1\times10^{-3}$ \cite{Hacker:2008}. Our absolute values are 30\% larger than predicted by VMD, but agree mostly within their errors. 

The LO values for the decay widths labeled LO agree within 20\% with the experimental values. The slopes are equal to zero at that order. At LO, taking the $1/N_c$ expansion of $Q$ into account leads to results that are far from the experimental values. The NLO calculations improve the description of the decay widths. In the NLO 1 case, the $\pi^0$ decay width is equal to the experimental value, because $L^{6,\epsilon}_8$ is fixed to it. In the NLO 2 case, where the parameters where fitted to all three decay widths, the description of $\Gamma_{\pi^0}$ worsens, while $\Gamma_\eta$ and $\Gamma_{\eta'}$ come closer to the experimental values. Our NLO 1-3 results for the slope of the $\eta$ agree well with the experimental value.
The description of the $\eta'$ slope, however, is very bad. Due to the small error of $b_\eta$ the fit favors this value, contributing to the poor description of $b_{\eta'}$. In the simultaneous fit to all decay widths and slope parameters (NLO 3), the results for the decay widths show larger deviations from the experimental values in comparison to NLO 2, marginally improving the values for the slopes.
In the NLO, Qexp scenario, the $\pi^0$ decay width is given by the leading-order value of the $Q(3)$ case. Since, then, the two parameters $L^{6,\epsilon}_8$ and $\lambda_1$ need to be fixed by $\Gamma_\eta$ and $\Gamma_{\eta'}$ alone, we reproduce the experimental values for these widths. The results for $b_\eta$ and $b_{\eta'}$ are very poor in this case. In the NLO, Qexp case, the errors of the LECs and the results for the decay widths and slopes are very large. This further reflects the fact that the NLO, Qexp calculation is not appropriate to describe the data, and the LECs cannot be fixed in a sensible way.
We thus conclude that omitting the $1/N_c$ expansion of $Q$ leads to a better description of the experimental data at LO and NLO. However, in general, the NLO calculation is not sufficient to adequately describe the decay widths and slopes of $\pi^0$, $\eta$, and $\eta'$, which motivates taking higher-order corrections into account.

\subsection{NNLO}
\label{sec:tpdNNLO}

\subsubsection{Parametrization of the TFFs and determination of the parameters}
At NNLO, a lot of new LECs appear both from the even-intrinsic-parity sector and the odd-intrinsic-parity sector. Moreover, our power counting demands taking terms of the $\mathcal{O}(p^8)$ Lagrangian into account, which has not been constructed yet. We therefore make the following ansatz for the $q^2$ dependence of the single-virtual TFFs up to and including NNLO:
\begin{align}
F_{\pi^0\gamma^\ast\gamma}(q^2)&=F_{\pi^0\gamma^\ast\gamma}^{\text{LO}}+\frac{1}{4\pi^2F_\pi}\left[A_{\pi^0}+B_{\pi^0} q^2+C_{\pi^0} {(q^2)}^2\right]+loops_{\pi^0}(q^2),\\
F_{\eta\gamma^\ast\gamma}(q^2)&=F_{\eta\gamma^\ast\gamma}^{\text{LO}}+\frac{1}{4\sqrt{3}\pi^2F_\pi}\left[A_\eta+B_\eta q^2+C_\eta {(q^2)}^2\right]+loops_\eta(q^2),\\
F_{\eta'\gamma^\ast\gamma}(q^2)&=F_{\eta'\gamma^\ast\gamma}^{\text{LO}}+\frac{2\sqrt{2}}{4\sqrt{3}\pi^2F_\pi}\left[A_{\eta'}+B_{\eta'} q^2+C_{\eta'} {(q^2)}^2\right]+loops_{\eta'}(q^2).
\end{align}
The $A_P$ and $B_P$ are combinations of LECs from the higher-order Lagrangians in Sec.~\ref{sec:Lagrangians} 
and, in principle, receive contributions from the $\mathcal{O}(p^8)$ Lagrangian as well. The $C_P$ stem solely from the $\mathcal{O}(p^8)$ Lagrangian. The expression $loops_P(q^2)$ denotes the $q^2$-dependent part of the loop corrections, while the $q^2$-independent parts are absorbed in the parameters $A_P$.
We determine the parameters $A_P$, $B_P$, $C_P$ through a simultaneous fit to the real-photon decay widths $\Gamma_{P\to\gamma\gamma}$ and to the experimental data for the TFFs. 
In the following, we perform several fits for the $\pi^0$, the $\eta$, and the $\eta'$ TFF considering different NNLO contributions. We start by fitting the full NNLO expressions. Then, we consider the case without loops, which means switching off the $q^2$-dependent pieces $loops_P(q^2)$. To study the influence of the $C_P$ terms, we also perform fits where we put $C_P=0$. Finally, we discuss the case where both $C_P$ and loops are neglected. In addition, we examine each of these four scenarios taking the $1/N_c$ expansion of $Q$ into account, denoted by Qexp. The fits are performed using the \textit{Mathematica} routine NonlinearModelFit, and the errors of the fit parameters are the ones obtained by this routine. A systematic error of at least 4\%, corresponding roughly to $\delta^3=(1/3)^3$,
should be added to all results determined up to NNLO. 

The $\pi^0$ TFF is fitted to both the time-like experimental data in Refs.~\cite{Achasov:2003ed,Akhmetshin:2004gw,Achasov:2016bfr,Adlarson:2016hpp} and the space-like data from Ref.~\cite{Danilkin:2019mhd}. For each scenario, we fit the TFF to four different regions of the photon virtuality $q^2$, which are given by $-0.55\ \text{GeV}^2\leq q^2 \leq 0.55\ \text{GeV}^2$, $-0.5\ \text{GeV}^2\leq q^2 \leq 0.5\ \text{GeV}^2$, $-0.45\ \text{GeV}^2\leq q^2 \leq 0.45\ \text{GeV}^2$, and $-0.4\ \text{GeV}^2\leq q^2 \leq 0.4\ \text{GeV}^2$. The results for the fit parameters obtained in the range $-0.5\ \text{GeV}^2\leq q^2 \leq 0.5\ \text{GeV}^2$ are provided in Table \ref{tab:pi0NNLO}, while the results for the other fits are shown in Table \ref{tab:pi0NNLOFull} in Appendix \ref{app:FitParameters}. 

The TFF of the $\eta$ is fitted to the time-like experimental data obtained in Refs.~\cite{Arnaldi:2009aa,Berghauser:2011zz,Aguar-Bartolome:2013vpw,Arnaldi:2016pzu,Adlarson:2016hpp}. For each case, we perform fits up to three different values of the invariant mass of the lepton pair, $m(l^+l^-)$. The maximal $m(l^+l^-)$ values are $m_1(l^+l^-)=0.47\ \text{GeV}$, $m_2(l^+l^-)=0.40\ \text{GeV}$, and $m_3(l^+l^-)=0.35\ \text{GeV}$. The results for the parameters fitted up to $0.47$ GeV are displayed in Table \ref{tab:etaNNLO}, and the results of the other fits can be found in Table \ref{tab:etaNNLOFull} in Appendix \ref{app:FitParameters}.

For the $\eta'$ TFF there are also data points in the space-like low-energy region available. Therefore, we fit the TFF to the space-like data from Ref.~\cite{Acciarri:1997yx} and to the time-like data from Ref.~\cite{Ablikim:2015wnx}. Here, we choose four fit regions for each scenario. The different fit ranges for the photon virtuality $q^2$ are $-0.53\ \text{GeV}^2\leq q^2 \leq 0.43\ \text{GeV}^2$ (I), $-0.53\ \text{GeV}^2\leq q^2 \leq 0.40\ \text{GeV}^2$ (II), $-0.50\ \text{GeV}^2\leq q^2 \leq 0.43\ \text{GeV}^2$ (III), and $-0.50\ \text{GeV}^2\leq q^2 \leq 0.40\ \text{GeV}^2$ (IV). Table \ref{tab:etapNNLO} shows the results for the parameters fitted in the range $-0.53\ \text{GeV}^2\leq q^2 \leq 0.43\ \text{GeV}^2$, and the results of the other fits are displayed in Table \ref{tab:etapNNLOFull} in Appendix \ref{app:FitParameters}. 

\begin{table}
	\centering
	\begin{tabular}{c r@{$\,\pm\,$}l r@{$\,\pm\,$}l r@{\,$\pm$\,}l}\myTR
		& \multicolumn{2}{c}{$A_{\pi^0}$} & \multicolumn{2}{c}{$B_{\pi^0}\ [\text{GeV}^{-2}]$} &  \multicolumn{2}{c}{$C_{\pi^0}\ [\text{GeV}^{-4}]$}\\ \myMR
		 \text{Full} & $-0.01$&$0.01$ & $0.91$&$0.15$ &
		$1.40$&$0.35$ \\
		\text{W/o loops} & $-0.01$&$0.01$ & $1.28$&$0.15$ &
		$1.62$&$0.35$ \\
		\text{$C_{\pi^0}=0$} & $-0.01$&$0.01$ & $0.34$&$0.06$ &
		$0$&$0$ \\
		\text{W/o loops $\wedge$ $C_{\pi^0}=0$} & $-0.01$&$0.01$ & $0.61$&$0.06$
		& $0$&$0$ \\
		\text{Full, Qexp} & $0.99$&$0.01$ & $1.25$&$0.15$ &
		$1.60$&$0.35$ \\
		\text{W/o loops, Qexp} & $0.99$&$0.01$ & $1.28$&$0.15$
		& $1.62$&$0.35$ \\
		\text{$C_{\pi^0}=0$, Qexp} & $0.99$&$0.01$ & $0.60$&$0.06$ &
		$0$&$0$ \\
		\text{W/o loops $\wedge$ $C_{\pi^0}=0$, Qexp} & $0.99$&$0.01$ &
		$0.61$&$0.06$ & $0$&$0$ \\
		\myBR
	\end{tabular}
	\caption{Fit parameters for the $\pi^0$ TFF. The LECs were fitted in the range $-0.5\ \text{GeV}^2\leq q^2 \leq 0.5\ \text{GeV}^2$.}
	\label{tab:pi0NNLO}
\end{table}

\begin{table}
	\centering
		\begin{tabular}{c r@{$\,\pm\,$}l r@{$\,\pm\,$}l r@{\,$\pm$\,}l}\myTR
	 & \multicolumn{2}{c}{$A_\eta$} & \multicolumn{2}{c}{$B_\eta\ [\text{GeV}^{-2}]$} &  \multicolumn{2}{c}{$C_\eta\ [\text{GeV}^{-4}]$}\\ \myMR
 \text{Full} & $-0.17$&$0.03$ & $2.32$&$0.22$ & $10.51$&$1.82$ \\
\text{W/o loops} & $-0.17$&$0.03$ & $2.98$&$0.22$ & $11.35$&$1.82$ \\
\text{$C_\eta=0$} & $-0.17$&$0.03$ & $3.41$&$0.12$ & $0$&$0$ \\
\text{W/o loops $\wedge$ $C_\eta=0$} & $-0.17$&$0.03$ & $4.16$&$0.12$ & $0$&$0$ \\
\text{Full, Qexp} & $0.66$&$0.03$ & $2.39$&$0.22$ & $10.59$&$1.82$\\
\text{W/o loops, Qexp} & $0.66$&$0.03$ & $2.98$&$0.22$ & $11.35$&$1.82$ \\
\text{$C_\eta=0$, Qexp} & $0.66$&$0.03$ & $3.49$&$0.12$ & $0$&$0$ \\
\text{W/o loops $\wedge$ $C_\eta=0$, Qexp} & $0.66$&$0.03$ & $4.16$&$0.12$ & $0$&$0$ \\
	\myBR
		\end{tabular}
		\caption{Fit parameters for the $\eta$ TFF. The LECs were fitted up to $0.47\ \text{GeV}$.}
		\label{tab:etaNNLO}
\end{table}

\begin{table}
	\centering
			\begin{tabular}{c r@{$\,\pm\,$}l r@{$\,\pm\,$}l r@{\,$\pm$\,}l}\myTR
	 & \multicolumn{2}{c}{$A_{\eta'}$} & \multicolumn{2}{c}{$B_{\eta'}\ [\text{GeV}^{-2}]$} &  \multicolumn{2}{c}{$C_{\eta'}\ [\text{GeV}^{-4}]$}\\ \myMR
 \text{Full} & $-0.06$&$0.02$ & $1.08$&$0.25$ & $1.18$&$0.52$ \\
\text{W/o loops} & $-0.06$&$0.02$ & $1.23$&$0.26$ & $1.30$&$0.52$ \\
\text{ $C_{\eta'}=0$} & $-0.06$&$0.02$ & $0.55$&$0.10$ & $0$&$0$ \\
\text{W/o loops $\wedge$ $C_{\eta'}=0$} & $-0.06$&$0.02$ & $0.64$&$0.11$ & $0$&$0$ \\
\text{Full, Qexp} & $-0.29$&$0.02$ & $1.07$&$0.25$ & $1.17$&$0.52$ \\
\text{W/o loops, Qexp} & $-0.29$&$0.02$ & $1.23$&$0.26$ & $1.3$&$0.52$ \\
\text{$C_{\eta'}=0$ , Qexp} & $-0.29$&$0.02$ & $0.54$&$0.10$ & $0$&$0$ \\
\text{W/o loops $\wedge$ $C_{\eta'}=0$, Qexp} & $-0.29$&$0.02$ & $0.64$&$0.11$ & $0$&$0$ \\
			\myBR
		\end{tabular}
		\caption{Fit parameters for the $\eta'$ TFF. The LECs were fitted in the range $-0.53\ \text{GeV}^2\leq q^2 \leq 0.43\ \text{GeV}^2$.}
		\label{tab:etapNNLO}
\end{table}

\begin{figure}[htbp]
	\centering
		\includegraphics[width=0.75\textwidth]{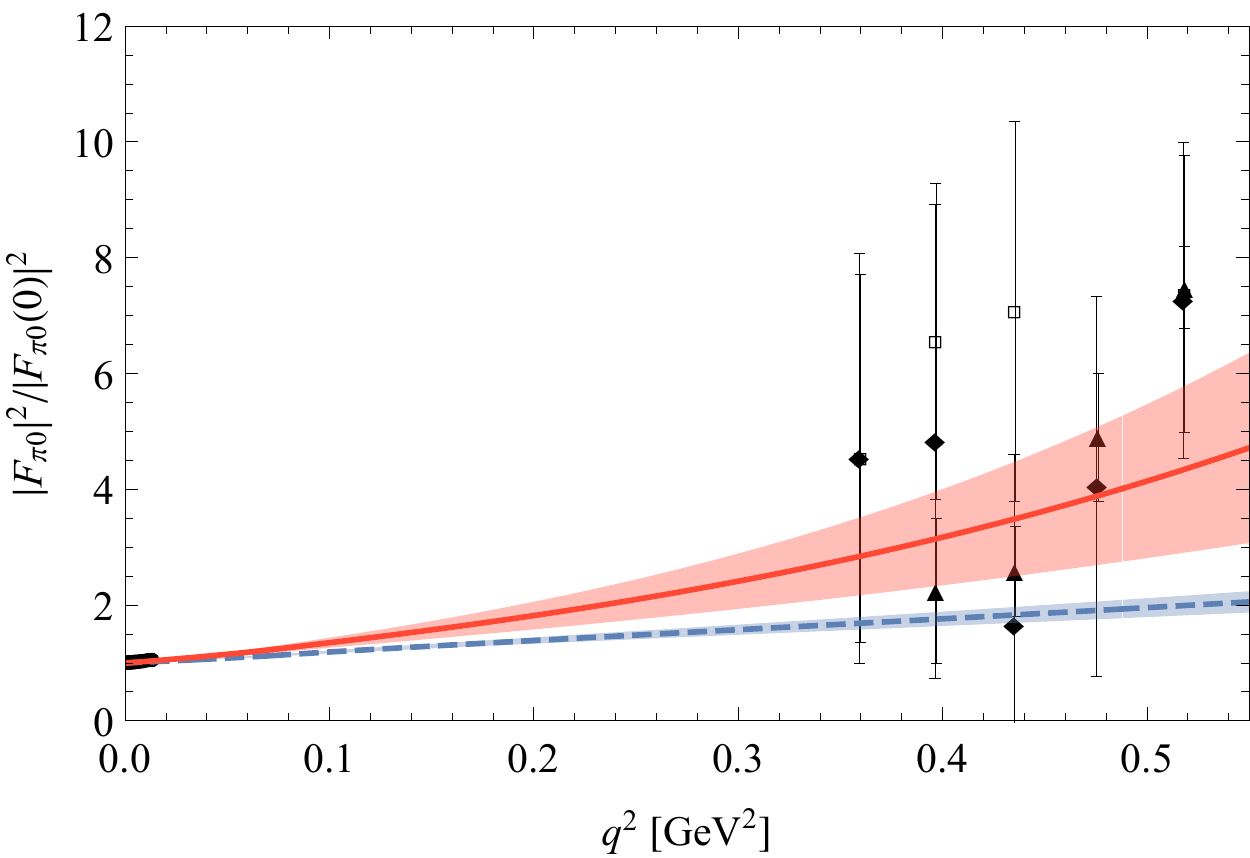}
	\caption{$\pi^0$ TFF in the time-like region, fitted in the range $-0.5\ \text{GeV}^2\leq q^2 \leq 0.5\ \text{GeV}^2$. The solid (red) line is the full NNLO calculation and the dashed (blue) line the NNLO result with $C_{\pi^0}=0$. The experimental data are taken from Refs.~\cite{Achasov:2003ed} ($\blacklozenge$), \cite{Akhmetshin:2004gw} ($\square$), \cite{Achasov:2016bfr} ($\blacktriangle$), \cite{Adlarson:2016hpp} ($\bullet$).}
	\label{fig:TFFbandspi0Tl}
\end{figure}

\begin{figure}[htbp]
	\centering
		\includegraphics[width=0.75\textwidth]{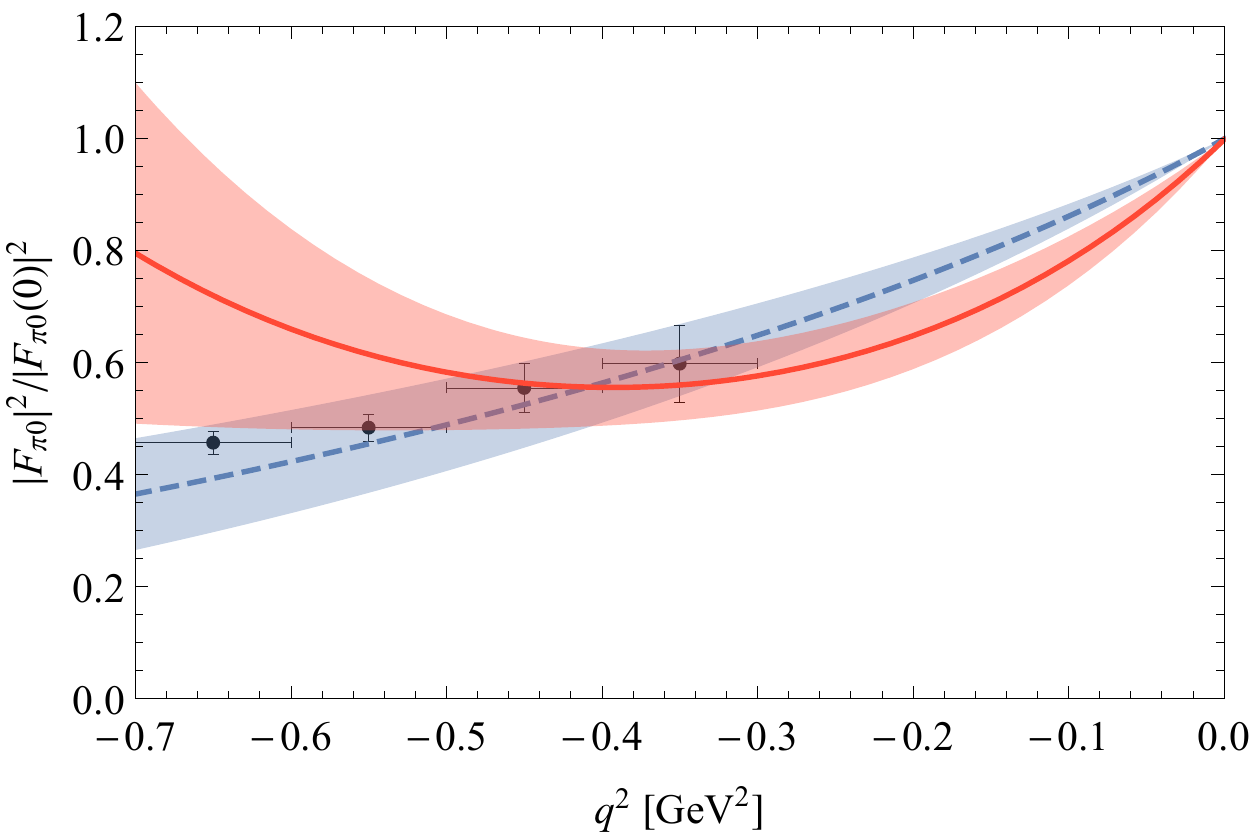}
	\caption{$\pi^0$ TFF in the space-like region, fitted in the range $-0.5\ \text{GeV}^2\leq q^2 \leq 0.5\ \text{GeV}^2$. The solid (red) line is the full NNLO calculation and the dashed (blue) line the NNLO result with $C_{\pi^0}=0$. The experimental
data are taken from Ref.~\cite{Danilkin:2019mhd}.}
	\label{fig:TFFbandspi0Sl}
\end{figure}

The $q^2$ dependence of the $\pi^0$ TFF is displayed in Fig.~\ref{fig:TFFbandspi0Tl} in the time-like region and in Fig.~\ref{fig:TFFbandspi0Sl} in the space-like region. Here, the TFF is normalized to $1$ at $q^2=0$ and plotted together with the experimental data. In this case, the TFF was fitted in the range $-0.5\ \text{GeV}^2\leq q^2 \leq 0.5\ \text{GeV}^2$. The bands show the $1\sigma$ error bands obtained by the \textit{Mathematica} fit routine NonlinearModelFit. 

\begin{figure}[htbp]
	\centering
		\includegraphics[width=0.75\textwidth]{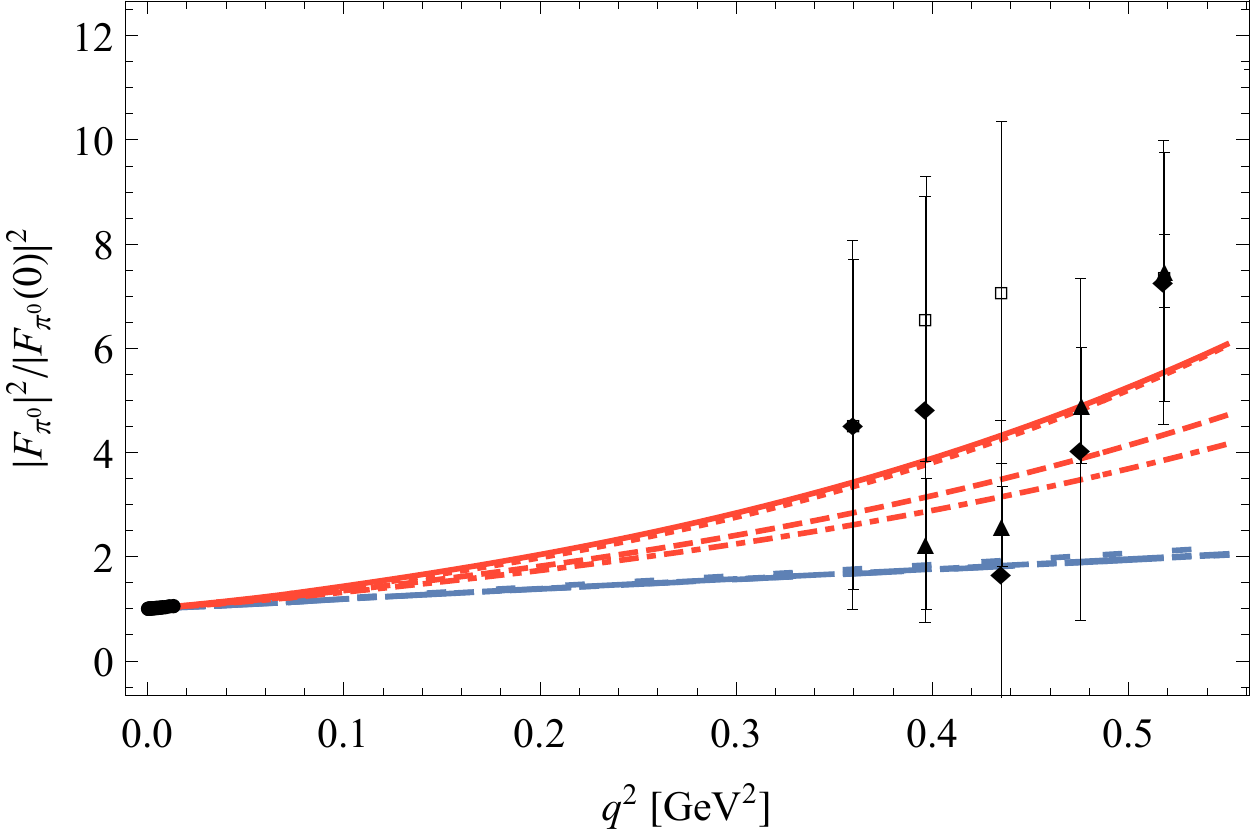}
	\caption{$\pi^0$ TFF in the time-like region fitted in the range $-0.55\ \text{GeV}^2\leq q^2 \leq 0.55\ \text{GeV}^2$ (solid, long-dashed), $-0.5\ \text{GeV}^2\leq q^2 \leq 0.5\ \text{GeV}^2$ (dashed, long-dash-dotted), $-0.45\ \text{GeV}^2\leq q^2 \leq 0.45\ \text{GeV}^2$ (dash-dotted, double-dotted), and $-0.4\ \text{GeV}^2\leq q^2 \leq 0.4\ \text{GeV}^2$ (dotted, long-dotted). The red (solid, dashed, dash-dotted, dotted) lines are the full NNLO calculations and the blue (long-dashed, long-dash-dotted, double-dotted, long-dotted) lines the NNLO results with $C_{\pi^0}=0$. The long-dashed, long-dash-dotted, and double-dotted curves are indistinguishable at this scale. The experimental
data are taken from Refs.~\cite{Achasov:2003ed} ($\blacklozenge$), \cite{Akhmetshin:2004gw} ($\square$), \cite{Achasov:2016bfr} ($\blacktriangle$), \cite{Adlarson:2016hpp} ($\bullet$).}
	\label{fig:TFFrangepi0Tl}
\end{figure}

\begin{figure}[htbp]
	\centering
		\includegraphics[width=0.75\textwidth]{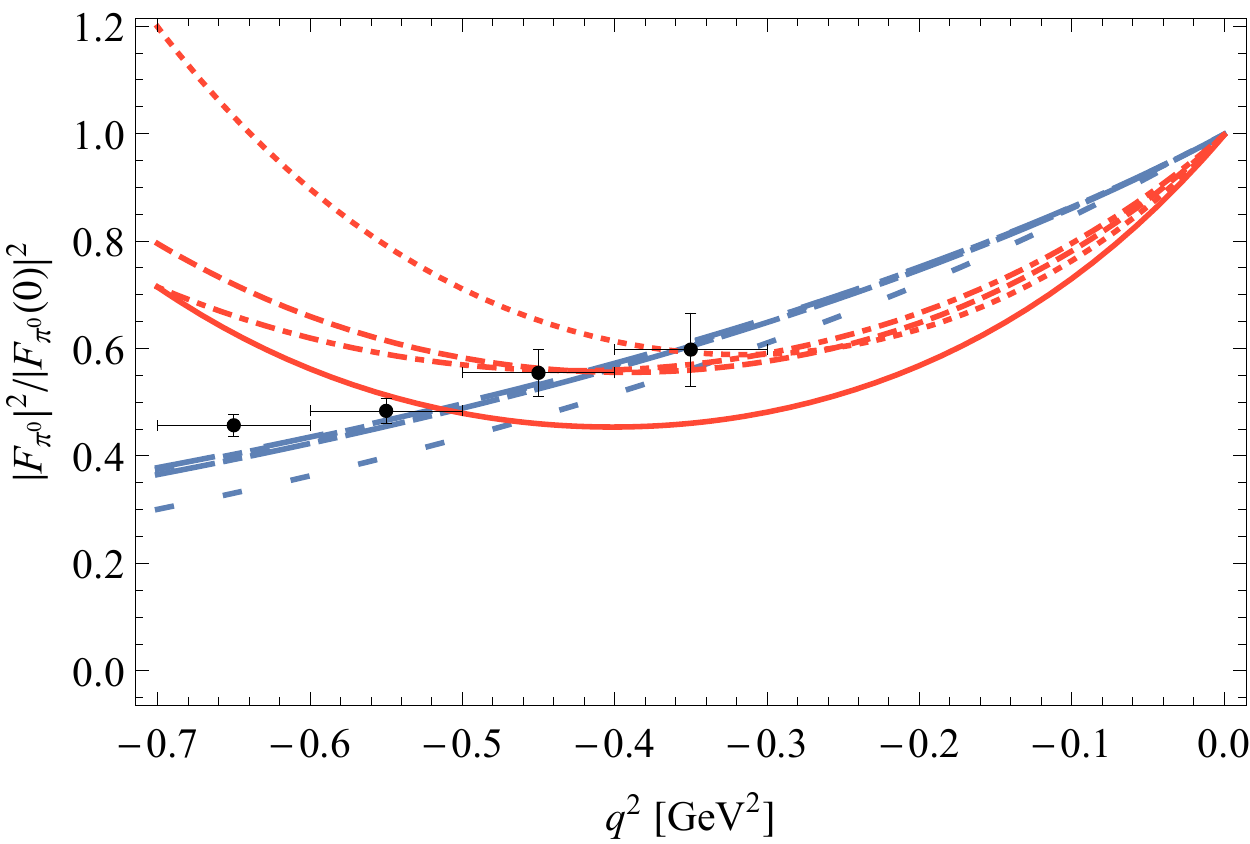}
	\caption{$\pi^0$ TFF in the space-like region fitted in the range $-0.55\ \text{GeV}^2\leq q^2 \leq 0.55\ \text{GeV}^2$ (solid, long-dashed), $-0.5\ \text{GeV}^2\leq q^2 \leq 0.5\ \text{GeV}^2$ (dashed, long-dash-dotted), $-0.45\ \text{GeV}^2\leq q^2 \leq 0.45\ \text{GeV}^2$ (dash-dotted, double-dotted), and $-0.4\ \text{GeV}^2\leq q^2 \leq 0.4\ \text{GeV}^2$ (dotted, long-dotted). The red (solid, dashed, dash-dotted, dotted) lines are the full NNLO calculations and the blue (long-dashed, long-dash-dotted, double-dotted, long-dotted) lines the NNLO results with $C_{\pi^0}=0$. The long-dashed, long-dash-dotted, and double-dotted curves are indistinguishable at this scale. The experimental
data are taken from Ref.~\cite{Danilkin:2019mhd}.}
	\label{fig:TFFrangespi0Sl}
\end{figure}

In Figs.~\ref{fig:TFFrangepi0Tl} and \ref{fig:TFFrangespi0Sl}, the results for the different fit ranges are displayed in the time-like and in the space-like region, respectively. 

\begin{figure}[htbp]
	\centering
		\includegraphics[width=0.9\textwidth]{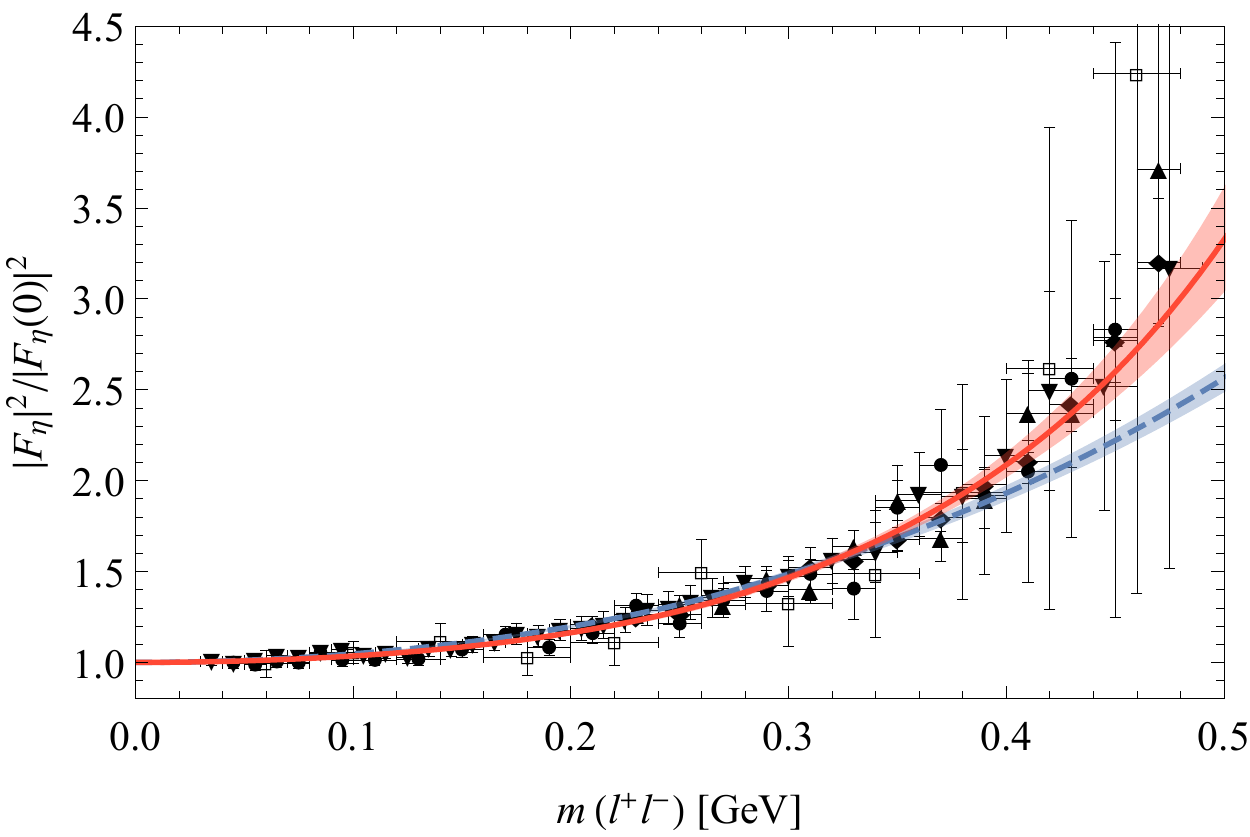}
	\caption{$\eta$ TFF fitted up to $0.47$ GeV. The solid (red) line is the full NNLO calculation and the dashed (blue) line the NNLO result with $C_\eta=0$. The experimental data are taken from Refs.~\cite{Adlarson:2016hpp} ($\blacktriangledown$), \cite{Arnaldi:2009aa} ($\blacktriangle$), \cite{Berghauser:2011zz} ($\square$), \cite{Aguar-Bartolome:2013vpw} ($\bullet$), \cite{Arnaldi:2016pzu} ($\blacklozenge$).}
	\label{fig:TFFbands}
\end{figure}

\begin{figure}[htbp]
	\centering
		\includegraphics[width=0.9\textwidth]{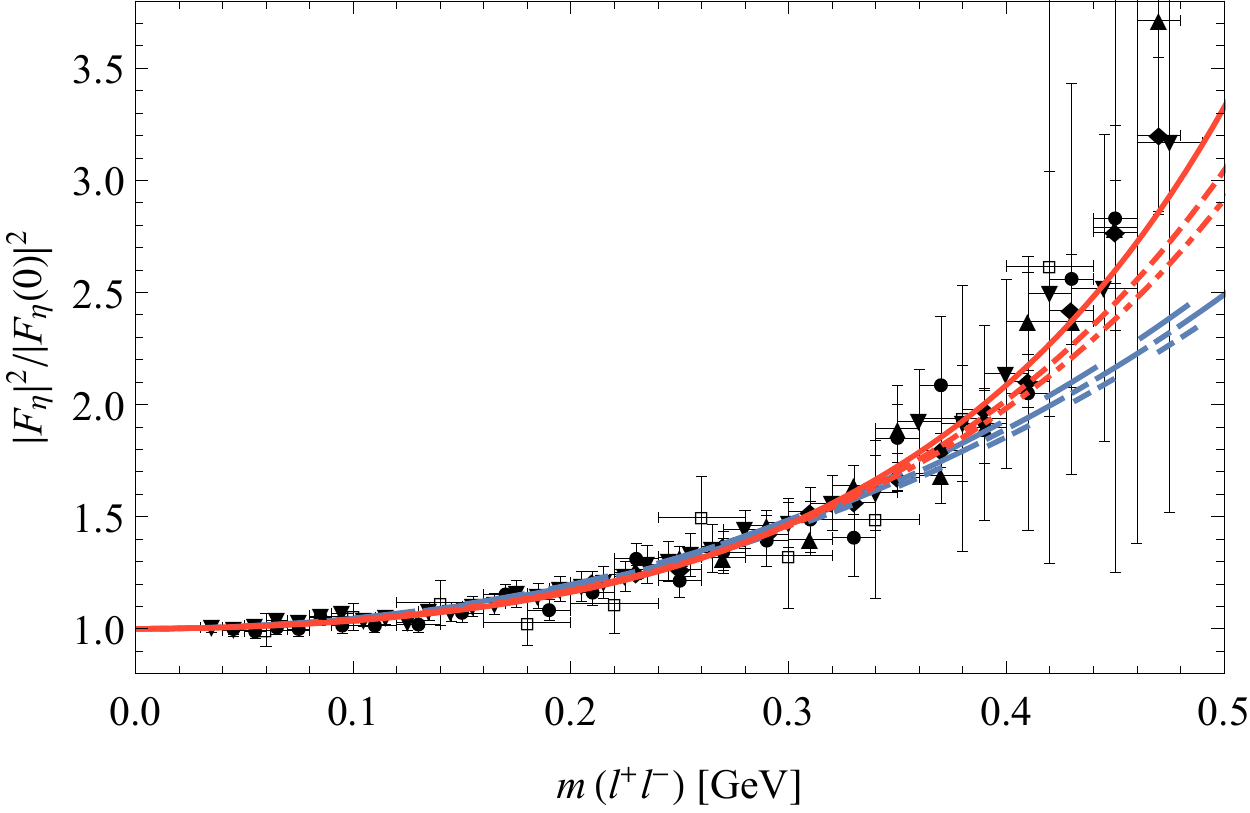}
	\caption{$\eta$ TFF fitted up to  $0.47$ GeV (solid, long-dashed), $0.40$ GeV (dashed, long-dash-dotted), and $0.35$ GeV (dash-dotted, double-dotted). The red (solid, dashed, dash-dotted) lines are the full NNLO calculations and the blue (long-dashed, long-dash-dotted, double-dotted) lines the NNLO results with $C_\eta=0$. The experimental data are taken from Refs.~\cite{Adlarson:2016hpp} ($\blacktriangledown$), \cite{Arnaldi:2009aa} ($\blacktriangle$), \cite{Berghauser:2011zz} ($\square$), \cite{Aguar-Bartolome:2013vpw} ($\bullet$), \cite{Arnaldi:2016pzu} ($\blacklozenge$)..}
	\label{fig:TFFrange}
\end{figure}

The $q^2$ dependence of the normalized $\eta$ TFF is shown in Fig.~\ref{fig:TFFbands}, where the TFF is plotted as a function of the invariant mass of the lepton pair $m(l^+l^-)$ together with the experimental data. In this case, the TFF was fitted up to $0.47$ GeV. The bands show the $1\sigma$ error bands. 
Figure \ref{fig:TFFrange} shows the results of the fits for the different fit ranges. As the fit range is extended to higher $m(l^+l^-)$ values, the curves become steeper. 

\begin{figure}[htbp]
	\centering
		\includegraphics[width=0.9\textwidth]{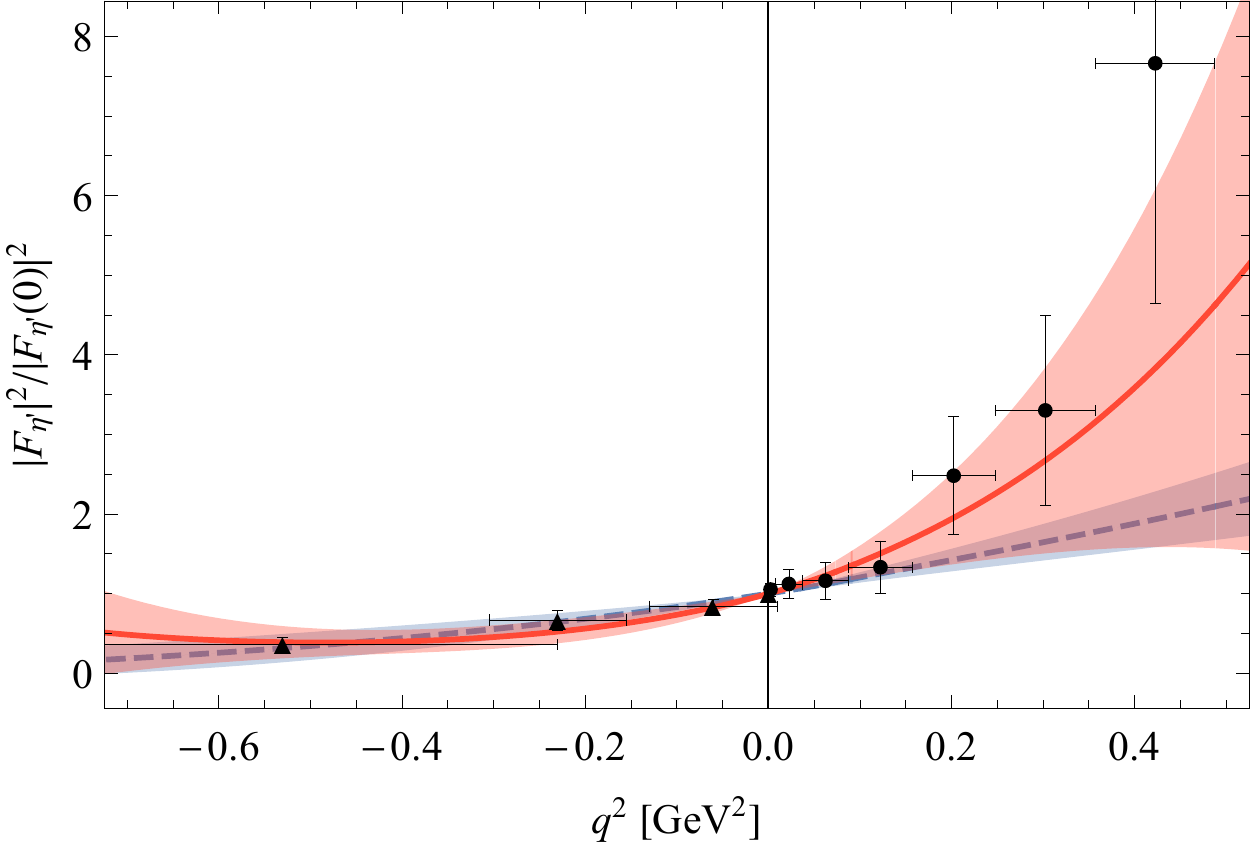}
	\caption{$\eta'$ TFF fitted in the range $-0.53\ \text{GeV}^2\leq q^2 \leq 0.43\ \text{GeV}^2$. The solid (red) line is the full NNLO calculation and the dashed (blue) line the NNLO result with $C_{\eta'}=0$. The time-like data are taken from Ref.~\cite{Ablikim:2015wnx} ($\bullet$) and the space-like data from Ref.~\cite{Acciarri:1997yx} ($\blacktriangle$).}
	\label{fig:TFFbandsp}
\end{figure}

\begin{figure}[htbp]
	\centering
		\includegraphics[width=0.9\textwidth]{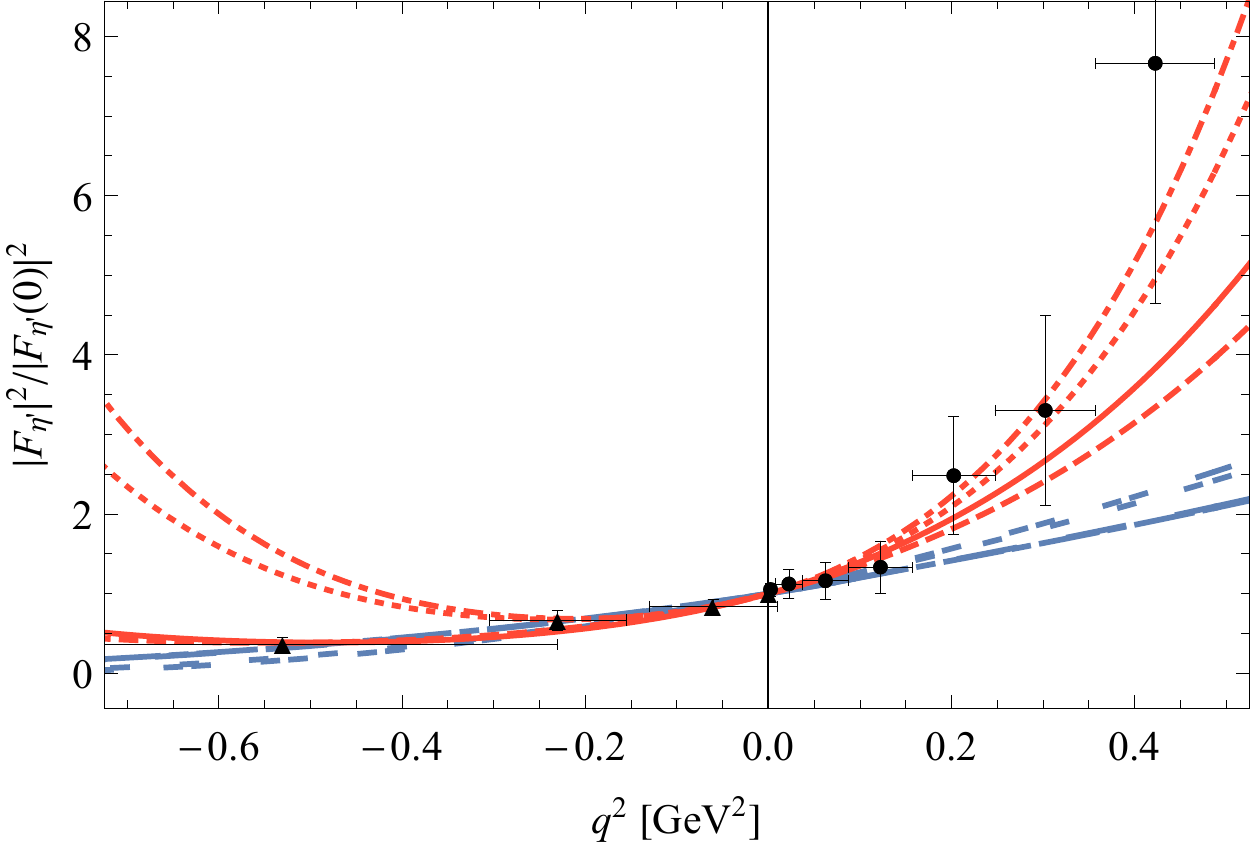}
	\caption{$\eta'$ TFF fitted in the range $-0.53\ \text{GeV}^2\leq q^2 \leq 0.43\ \text{GeV}^2$ (solid, long-dashed), $-0.53\ \text{GeV}^2\leq q^2 \leq 0.40\ \text{GeV}^2$ (dashed, long-dash-dotted), $-0.50\ \text{GeV}^2\leq q^2 \leq 0.43\ \text{GeV}^2$ (dash-dotted, double-dotted), $-0.50\ \text{GeV}^2\leq q^2 \leq 0.40\ \text{GeV}^2$
 (dotted, long-dotted). The red (solid, dashed, dash-dotted, dotted) lines are the full NNLO calculations and the blue (long-dashed, long-dash-dotted, double-dotted, long-dotted) lines the NNLO results with $C_{\eta'}=0$. The long-dashed and long-dash-dotted curves are indistinguishable at this scale. The time-like data are taken from Ref.~\cite{Ablikim:2015wnx} ($\bullet$) and the space-like data from Ref.~\cite{Acciarri:1997yx} ($\blacktriangle$).}
	\label{fig:TFFrangep}
\end{figure}

The $q^2$ dependence of the normalized $\eta'$ TFF, fitted between $-0.53\ \text{GeV}^2$ and $0.43\ \text{GeV}^2$, is shown in Fig.~\ref{fig:TFFbandsp} together with the experimental data. The bands are the 1$\sigma$ error bands due to the errors of the fit parameters. 
The results for the $\eta'$ TFF fitted to different ranges are displayed in Fig.~\ref{fig:TFFrangep}.

\subsubsection{Discussion of the results}

In the following, we will interpret the results of the NNLO analysis. We start with the discussion of the LECs determined at NNLO, shown in Tables \ref{tab:pi0NNLO}-\ref{tab:etapNNLO}.
Switching on or off the loop contributions corresponds to keeping the $q^2$-dependent parts, $loops_P(q^2)$, or neglecting them, respectively. As a result, the parameters $A_P$ remain the same in both cases. 
The inclusion of the $1/N_c$ expansion of $Q$ has almost no visible effect on the shape of the TFFs. However, this expansion has an influence on the parameters $A_P$, i.e., the absolute normalizations of the form factors, which change notably, since the LO expressions for the TFF (see Sec.~\ref{sec:TPDCalcA}) are different with or without the $1/N_c$ expansion of $Q$. The $q^2$-dependent loop corrections $loops_P(q^2)$ 
give numerically quite similar contributions to the TFF with or without the $1/N_c$ expansion of $Q$. Therefore, the parameters $B_P$ and $C_P$ do not vary very much in these two cases. 
As Figures \ref{fig:TFFcasespi0Tl} - \ref{fig:TFFcasesp} in Appendix \ref{app:plots} show, the influence of the loop contributions on the shape of the TFFs is very small. However, the effects of the loops can be seen in the variation of $B_P$ and $C_P$ with and without loops, where we observe rather small changes. Neglecting the loops leads to an increase of the values of $B_P$ and $C_P$ in order to compensate for the missing contributions, which add positively to the TFFs.

Table \ref{tab:pi0NNLO} in Appendix \ref{app:FitParameters} shows the variation of the fit parameters for the $\pi^0$ TFF with decreasing fit range. For $C_\pi^0$ equal to zero, the parameter $B_\pi^0$ increases only slightly if the fit range is decreased. In general, since the $B_\pi^0$ term is not able to provide curvature, the curves are rather flat. If the $C_\pi^0$ term is included, $B_\pi^0$ and $C_\pi^0$ decrease for the first three decreasing ranges, but increase for the last (smallest) range. When fewer data points are included, the curves become less steep, matching the decreasing parameter values. In the last scenario (IV), however, the fit is dominated by the single data point in the space-like region, leading to more curvature and therefore larger parameter values.

The results for $\eta$ TFF fit parameters are displayed in Table \ref{tab:etaNNLOFull} in Appendix \ref{app:FitParameters} for decreasing fit ranges. If $C_\eta$ is put to zero, the parameter $B_\eta$ decreases as the fit range decreases. This behavior is in accordance with the fact that the curves become steeper as the fit range is extended to higher $q^2$ values. If we include the $C_\eta$ term in the fit, there is an interplay between $C_\eta$ and $B_\eta$. For decreasing fit range, the $C_\eta$ values tend to decrease while $B_\eta$ increases. In addition, the errors of $B_\eta$ and $C_\eta$ become larger. This is to be expected, since less data points are included in the fit, there seems to be a correlation between $B_\eta$ and $C_\eta$, and the $C_\eta$ term becomes more important at higher values of $q^2$.

In the case of the $\eta'$, the fit range is varied both in the time-like and the space-like region. The variation of the parameters is displayed in Table \ref{tab:etapNNLOFull} in Appendix \ref{app:FitParameters}. Decreasing the time-like fit range yields smaller values for both $B_{\eta'}$ and $C_{\eta'}$. This is to be expected, since the TFF curves show less curvature as the fit range gets smaller.
If we exclude the last space-like data point, the values for $B_{\eta'}$ and $C_{\eta'}$ increase. In this case, the fit focuses more on the time-like region and the parameters adjust to the steep rise of the time-like TFF.

\subsubsection{Slope and curvature}

Employing the results for the fit parameters, we calculate the slopes and the curvatures of the TFFs as defined in Eqs.~(\ref{eq:slope}) and (\ref{eq:curv}). The errors are due to the errors of the fit parameters. As a first estimate, we assume that the fit parameters are independent. Taking into account their correlations is beyond the scope of this work. The main results are given in Tables \ref{tab:SlopeCurvPi}-\ref{tab:SlopeCurvp}.
\begin{table}[tbp]
	\centering
	\small{
		\begin{tabular}{c  c r@{$\,\pm\,$}l r@{$\,\pm\,$}l }\hline\hline%\toprule\toprule
 & $|q^2|_\text{max}$ [GeV$^2$]& \multicolumn{2}{c}{$b_{\pi^0}$} &  \multicolumn{2}{c}{$c_{\pi^0}$}  \\ \hline%\midrule
\text{Full} & 0.55 & 0.031 & 0.002 & 0.0008 & 0.0001 \\
 \text{Full} & 0.5 & 0.025 & 0.003 & 0.0007 & 0.0001 \\
 \text{Full} & 0.45 & 0.023 & 0.003 & 0.0006 & 0.0001 \\
 \text{Full} & 0.4 & 0.028 & 0.005 & 0.0009 & 0.0003 \\
 \text{$C_{\pi^0}=0$} & 0.55 & 0.014 & 0.001 & 0.0002 & 0.0000 \\
 \text{$C_{\pi^0}=0$} & 0.5 & 0.014 & 0.001 & 0.0002 & 0.0000 \\
 \text{$C_{\pi^0}=0$} & 0.45 & 0.014 & 0.001 & 0.0002 & 0.0000 \\
 \text{$C_{\pi^0}=0$} & 0.4 & 0.016 & 0.002 & 0.0002 & 0.0000 \\
\hline\hline%\bottomrule\bottomrule	
		\end{tabular}}
		\caption{Results for the slope and the curvature of the $\pi^0$ TFF at NNLO, fitted up to $|q^2|_\text{max}$.}
		\label{tab:SlopeCurvPi}
		\end{table}
		
\begin{table}[tbp]
	\centering
	\small{
		\begin{tabular}{c c r@{$\,\pm\,$}l r@{$\,\pm\,$}l }\myTR
 & $m_{\max}(l^+l^-)$ [GeV] & \multicolumn{2}{c}{$b_{\eta}$} &  \multicolumn{2}{c}{$c_{\eta}$}  \\ \myMR
 \text{Full} & 0.47 & 0.51 & 0.04 & 0.60 & 0.10 \\
\text{Full} & 0.4 & 0.55 & 0.05 & 0.43 & 0.10 \\
\text{Full} & 0.35 & 0.57 & 0.06 & 0.36 & 0.10 \\
\text{$C_\eta=0$} & 0.47 & 0.70 & 0.02 & 0.05 & 0.00 \\
\text{$C_\eta=0$} & 0.4 & 0.67 & 0.02 & 0.05 & 0.00 \\
\text{$C_\eta=0$} & 0.35 & 0.65 & 0.03 & 0.05 & 0.00 \\
\myBR
		\end{tabular}}
		\caption{Results for the slope and the curvature of the $\eta$ TFF at NNLO.}
		\label{tab:SlopeCurv}
		\end{table}
		
	\begin{table}[tbp]
	\centering
	\small{
		\begin{tabular}{c c r@{$\,\pm\,$}l r@{$\,\pm\,$}l }\myTR
 & Fit range & \multicolumn{2}{c}{$b_{\eta'}$} &  \multicolumn{2}{c}{$c_{\eta'}$}  \\ \myMR
 \text{Full} & \text{I} & 1.47 & 0.31 & 1.58 & 0.57 \\
\text{Full} & \text{II} & 1.32 & 0.33 & 1.30 & 0.57 \\
\text{Full} & \text{III} & 1.52 & 0.28 & 3.46 & 0.58 \\
\text{Full} & \text{IV} & 1.42 & 0.31 & 2.95 & 0.58 \\
\text{$C_{\eta'}=0$} & \text{I} & 0.84 & 0.13 & 0.28 & 0.01 \\
\text{$C_{\eta'}=0$} & \text{II} & 0.82 & 0.13 & 0.28 & 0.01 \\
\text{$C_{\eta'}=0$} & \text{III} & 1.11 & 0.27 & 0.28 & 0.01 \\
\text{$C_{\eta'}=0$} & \text{IV} & 1.04 & 0.27 & 0.28 & 0.01 \\
\myBR	
		\end{tabular}}
		\caption{Results for the slope and the curvature of the $\eta'$ TFF at NNLO. The fit ranges are: $-0.53\ \text{GeV}^2\leq q^2 \leq 0.43\ \text{GeV}^2$ (I), $-0.53\ \text{GeV}^2\leq q^2 \leq 0.40\ \text{GeV}^2$ (II), $-0.50\ \text{GeV}^2\leq q^2 \leq 0.43\ \text{GeV}^2$ (III), $-0.50\ \text{GeV}^2\leq q^2 \leq 0.40\ \text{GeV}^2$ (IV).}
		\label{tab:SlopeCurvp}
		\end{table}
		
The values for the slopes with and without loop contributions agree within their uncertainties. This is the case, because the influence of the loops is already compensated by different values for the fit parameters $B_P$. The $1/N_c$ expansion of the quark-charge matrix plays a negligible role. For $C_{\pi^0}=0$, the $\pi^0$ slope is smaller than in the full NNLO calculation. This is in accordance with the fact that the curves are less steep for $C_{\pi^0}=0$, since the fit is dominated by the values at very low $q^2$.  
If we neglect the $C_\eta$ term, $B_\eta$ compensates for the missing contribution, and, as a result, the $\eta$ slope increases. This effect is diminished, if the fit range is restricted to lower $q^2$ values. In the case of the $\eta'$, if we set $C_{\eta'}=0$, the slope gets smaller. This behavior is different from the one of the $\eta$ slope due to the inclusion of the space-like data. As a further check, we have investigated the case where the fit is performed only to the time-like data. Then, the $\eta'$ slope increases if we put $C_{\eta'}=0$, which is similar to the $\eta$ case.

\begin{figure}[htbp]
	\centering
	\includegraphics[width=1\textwidth]{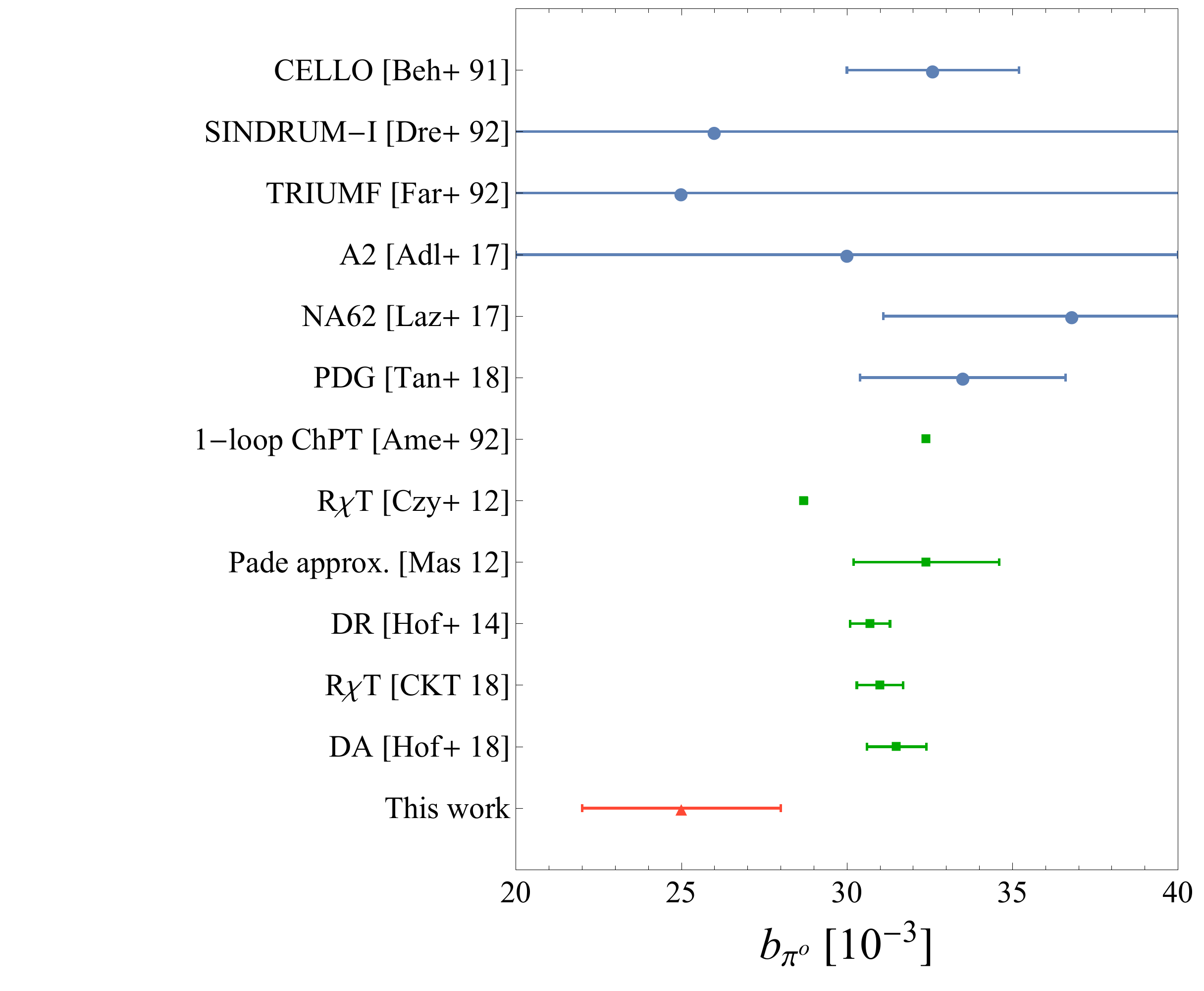}
	\caption{Our result for $b_{\pi^0}$ compared to other experimental and theoretical determinations from Beh+ 91 \cite{Behrend:1990sr}, Dre+ 92 \cite{MeijerDrees:1992qb}, Far+ 92 \cite{Farzanpay:1992pz}, Adl+ 17 \cite{Adlarson:2016ykr}, Laz+ 17 \cite{TheNA62:2016fhr}, Tan+ 18 \cite{Tanabashi:2018oca}, Ame+ 92 \cite{Ametller:1991jv}, Czy+ 12 \cite{Czyz:2012nq}, Mas 12 \cite{Masjuan:2012wy}, Hof+ 14 \cite{Hoferichter:2014vra}, CKT 18 \cite{Czyz:2017veo}, Hof+ 18 \cite{Hoferichter:2018kwz}.}
	\label{fig:splotpi0}
\end{figure}

\begin{figure}[htbp]
	\centering
		\includegraphics[width=1\textwidth]{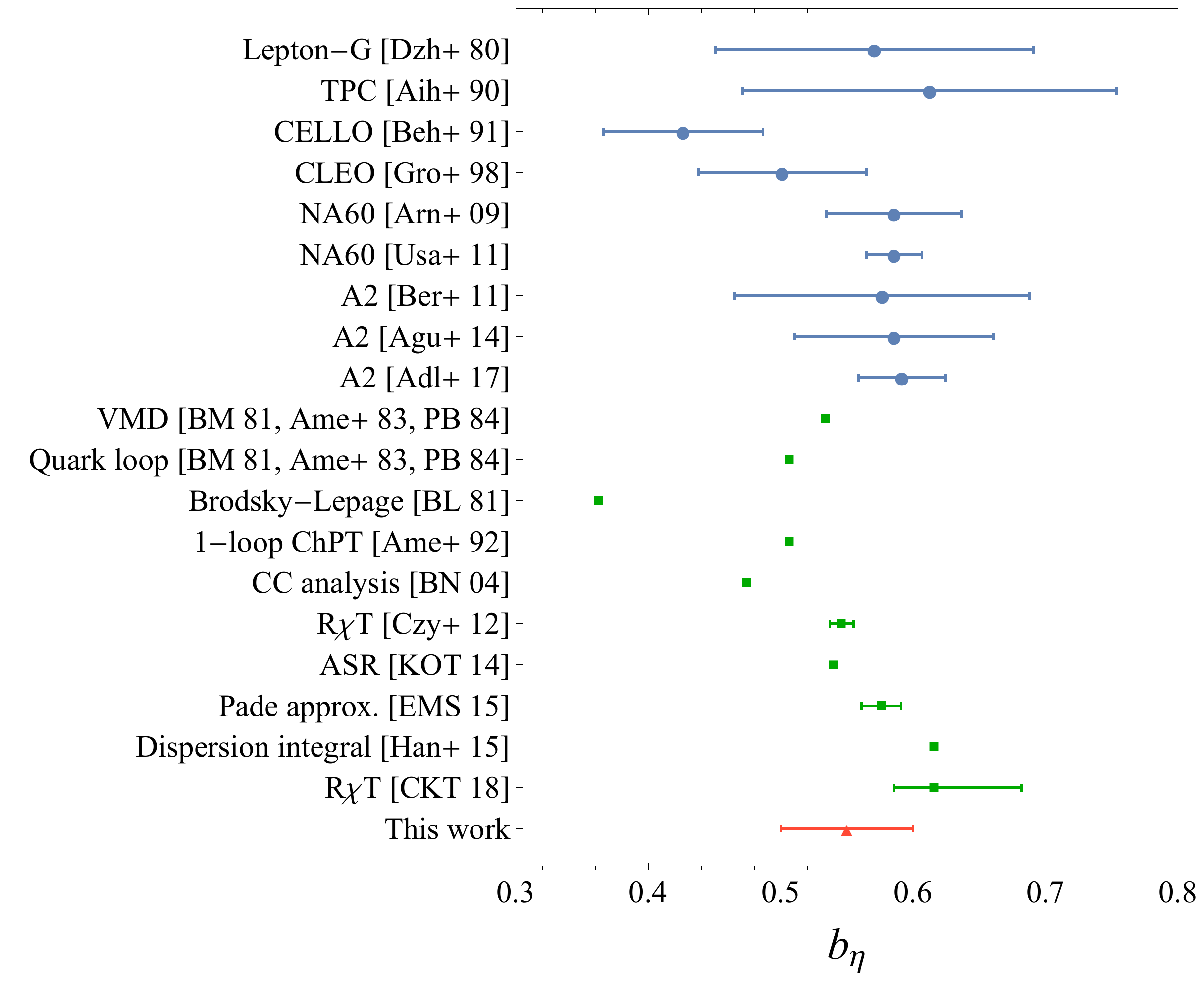}
	\caption{Our result for $b_\eta$ compared to other experimental and theoretical determinations from Dzh+ 80 \cite{Dzhelyadin:1980kh}, Aih+ 90 \cite{Aihara:1990nd}, Beh+ 91 \cite{Behrend:1990sr}, Gro+ 98 \cite{Gronberg:1997fj}, Arn+ 09 \cite{Arnaldi:2009aa}, Usa+ 11 \cite{Usai:2011zza}, Ber+ 11 \cite{Berghauser:2011zz}, Agu+ 14 \cite{Aguar-Bartolome:2013vpw}, Adl+ 17, BM 81 \cite{Bramon:1981sw}, Ame+ 83 \cite{Ametller:1983ec}, PB 84 \cite{Pich:1983zk}, BL 81 \cite{Brodsky:1981rp}, Ame+ 92 \cite{Ametller:1991jv}, BN 04 \cite{Borasoy:2003yb}, Czy+ 12 \cite{Czyz:2012nq}, KOT 14 \cite{Klopot:2013laa}, EMS 15 \cite{Escribano:2015nra}, Han+ 15 \cite{Hanhart:2013vba}, CKT 18 \cite{Czyz:2017veo}.}
	\label{fig:splot}
\end{figure}

\begin{figure}[htbp]
	\centering
		\includegraphics[width=1\textwidth]{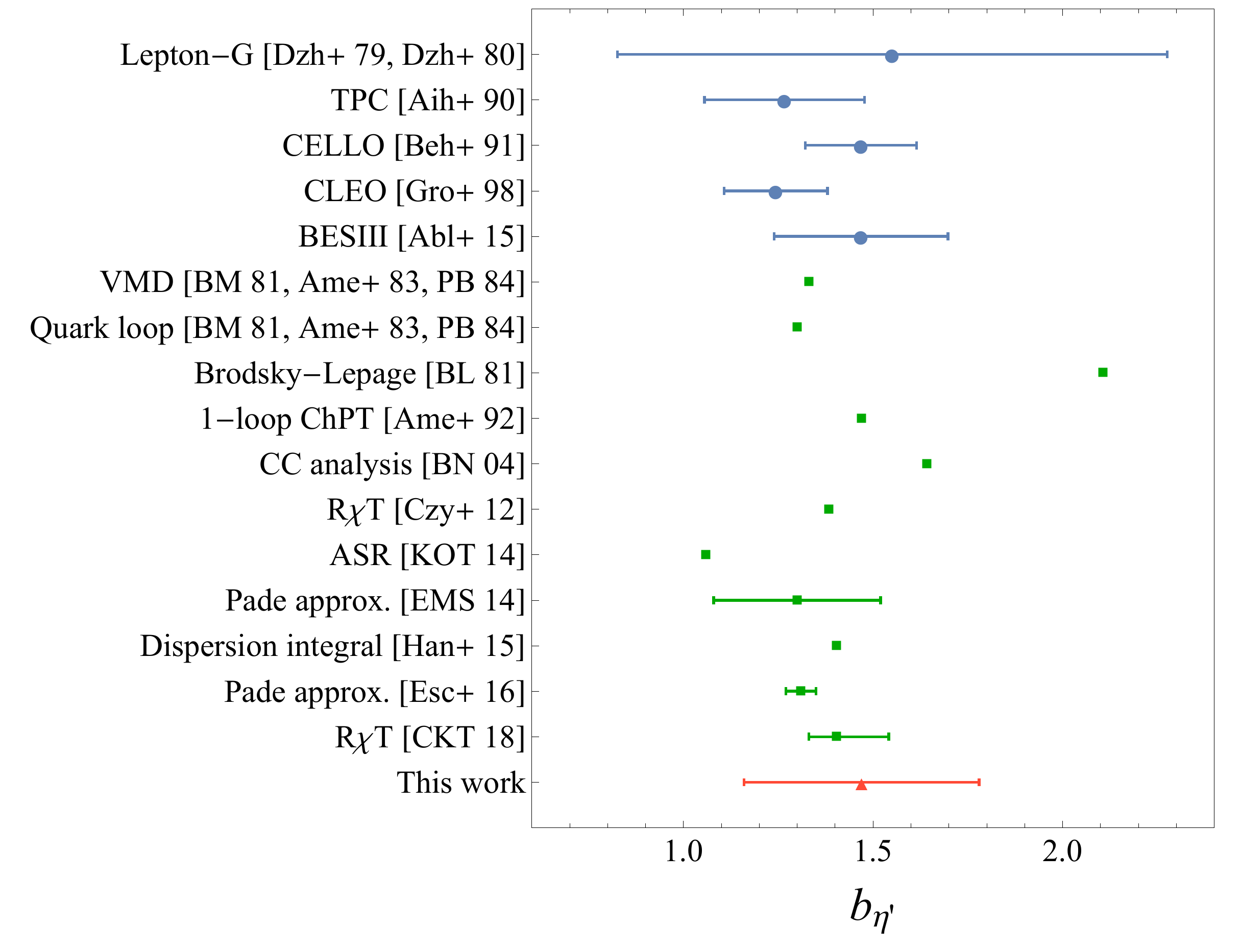}
	\caption{Our result for $b_{\eta'}$ compared to other experimental and theoretical determinations from Dzh+ 79 \cite{Dzhelyadin:1979za}, Dzh+ 80 \cite{Dzhelyadin:1980kh}, Aih+ 90 \cite{Aihara:1990nd}, Beh+ 91 \cite{Behrend:1990sr}, Gro+ 98 \cite{Gronberg:1997fj}, Abl+ 15 \cite{Ablikim:2015wnx}, BM 81 \cite{Bramon:1981sw}, Ame+ 83 \cite{Ametller:1983ec}, PB 84 \cite{Pich:1983zk}, BL 81 \cite{Brodsky:1981rp}, Ame+ 92 \cite{Ametller:1991jv}, BN 04 \cite{Borasoy:2003yb}, Czy+ 12 \cite{Czyz:2012nq}, KOT 14 \cite{Klopot:2013laa}, EMS 14 \cite{Escribano:2013kba}, Han+ 15 \cite{Hanhart:2013vba}, Esc+ 16 \cite{Escribano:2015yup}, CKT 18 \cite{Czyz:2017veo}.}
	\label{fig:splotp}
\end{figure}

Figures \ref{fig:splotpi0} - \ref{fig:splotp} show the comparison of our results for the $\pi^0$, $\eta$, and $\eta'$ slopes together with other experimental and theoretical determinations. 
Our values $b_\eta=0.55(5)$ from the fit up to $0.4\ \text{GeV}$, and $b_{\eta'}=1.47(31)$ from fit I agree within the errors with most of the other theoretical and experimental results. 
The result $b_{\pi^0}=0.025(3)$ from the fit up to $0.5\ \text{GeV}^2$, however, is smaller than the other theoretical predictions and most of the experimental results. This is due to the inclusion of higher-$q^2$ data. Our result for the fit only up to $0.5\ \text{GeV}$, $b_{\pi^0}=0.028(5)$, e.g., is closer to the other predictions.
In general, our result for $b_\eta$ is slightly lower than the other determinations, whereas our value for $b_{\eta'}$ is slightly higher than the other results.
From Table \ref{tab:SlopeCurv} one can observe that a decreasing fit range leads to values for $b_\eta$ which come closer to the other theoretical and experimental determinations. 

The main results for the curvatures of $\pi^0$, $\eta$, and $\eta'$ are displayed in Tables \ref{tab:SlopeCurvPi} - \ref{tab:SlopeCurvp}. As the slope, the $\pi^0$ curvature decreases with decreasing fit range except for the smallest range, where the curvature takes its largest value. In this case, the TFF is adjusted to the single data point in the space-like region leading to a large curvature. 
The $\eta$ curvature is reduced if the fit range is restricted to smaller $q^2$ values. The $\eta'$ curvature decreases if the time-like range is decreased. As the space-like fit range becomes smaller, the fit is dominated by the steeply rising time-like data and the curvature is almost twice as large.  
The main contributions to the curvature stem from the $C_{\eta'}$ terms. If we put them to zero, the remaining curvature is given by the loop contributions, which is rather small.

Our values for the curvatures can be compared with other theoretical determinations. A dispersive analysis finds $c_{\pi^0}=1.14(4)\times 10^{-3}$ \cite{Hoferichter:2018kwz}, and works using Pad\'{e} approximants obtain $c_{\pi^0}=1.06(26)\times 10^{-3}$ \cite{Masjuan:2012wy},  $c_\eta=0.339(15)_{\text{stat}}(5)_{\text{sys}}$ \cite{Escribano:2015nra}, and $c_{\eta'}=1.72(47)_{\text{stat}}(34)_{\text{sys}}$ \cite{Escribano:2013kba}. If we use a simple VMD estimate as given in Eq.~(\ref{eq:slopecurvVMD}) with $\Lambda_P=0.77\ \text{GeV}$ \cite{Hacker:2008}, we obtain $c_\pi^0=0.9\times 10^{-3}$, $c_\eta=0.26$, and $c_{\eta'}=2.40$.
Our results for $c_{\pi^0}$ are slightly smaller than the other predictions, being closest to the VMD one. For $c_\eta$, our values are mostly larger than the other predictions. Only if the fit range is decreased, our results come to agreement with Ref.~\cite{Escribano:2015nra}, whereas the naive VMD prediction is even smaller. In the cases including the full space-like data, the $\eta'$ curvature is slightly smaller than the one from Ref.~\cite{Escribano:2013kba}, but shows agreement within the errors. The VMD prediction for $c_{\eta'}$ is larger and lies on the upper end of the error band in Ref.~\cite{Escribano:2013kba}. None of our values reaches the VMD value within the error range. Note that the errors are only the ones provided by the fit.
The results for $c_{\eta'}$ in the cases where the space-like fit range is restricted are much larger than the ones from the fit to all space-like data as well as the ones from the other references.

\section{Single Dalitz decays}

Having performed the numerical evaluation of the single-virtual TFFs of $\pi^0$, $\eta$, and $\eta'$, we are now able to calculate the widths of the decays to one photon and a lepton pair. 
To obtain the invariant amplitude for the decay $P\to\gamma l^+l^-$, where $P=\pi^0,\ \eta,\ \eta'$ and $l=e,\ \mu$, we use Eq.~(\ref{matrixeTPD}) and therein define 
$q^\mu_1$, with $q^2_1=s$, and $\epsilon^\mu_1=(e/s)[\bar{u}\gamma^\mu v]$ as virtual-photon momentum and polarization, respectively. The momentum of the real photon is denoted by $q_2$ with $q^2_2=0$, and $\epsilon_2$ is its polarization.
The decay width can be written as \cite{Hacker:2008}
\begin{align}
\Gamma(P\to\gamma l^+l^-)=\int^{M^2_P}_{4m^2_l}ds\frac{\sqrt{\lambda[M^2_P,s,0]}\sqrt{\lambda[s,m^2_l,m^2_l]}}{1024M^3_P\pi^4s}\int d\Omega_{ll}\overline{\sum}|\mathcal{M}|^2.
\end{align}
Defining the leptonic tensor
\begin{align}
L^{\mu\nu}=\sum_{\text{spin}}[\bar{u}\gamma^\mu v][\bar{u}\gamma^\nu v]^*,
\end{align}
and employing the identity
\begin{align}
\int\frac{d\Omega_{ll}}{4\pi}L_{\mu\mu'}=\frac{4}{3}\left(1+\frac{2m^2_l}{q^2}\right)(q_\mu q_{\mu'}-q^2g_{\mu\mu'}),\quad q^2=s,
\end{align}
one obtains
\begin{align}
&\Gamma(P\to\gamma l^+l^-)\no\\
&=\frac{e^2}{384M^3_P\pi^3}\int^{M^2_P}_{4m^2_l}ds\frac{\sqrt{1-\frac{4m^2_l}{s}}(M^2_P-s)^3(2m^2_l+s)}{s^2}\left|F_{P\gamma l^+l^-}\right|^2.
\end{align}
To evaluate this expression numerically, we make use of the LECs determined in Secs. \ref{sec:tpdNLO} and \ref{sec:tpdNNLO}. At NNLO, we employ the LECs determined from the fit of the $\pi^0$ TFF up to $0.5\ \text{GeV}^2$, the $\eta$ TFF up to $0.47\ \text{GeV}$, and the $\eta'$ TFF in the range $-0.53\ \text{GeV}^2\leq q^2 \leq 0.43\ \text{GeV}^2$ (I). 
The results for the decay widths to one photon and a lepton pair are shown in Tables \ref{tab:ee} and \ref{tab:mumu}. The errors are calculated from the errors of the LECs which are assumed to be uncorrelated. They can be viewed as upper limits for the errors. Taking the correlations into account is beyond the scope of this work. 

\begin{table}[tbp]
	\centering
		\begin{tabular}{c  r@{$\,\pm\,$}l r@{$\,\pm\,$}l r@{$\,\pm\,$}l r@{$\,\pm\,$}l r@{$\,\pm\,$}l r@{$\,\pm\,$}l }\myTR
			& \multicolumn{2}{c}{$\Gamma_{\pi^0\to\gamma e^+ e^-}$} &  \multicolumn{2}{c}{$\text{BR}^{\text{rel}}_{\pi^0\to\gamma e^+e^-}$}
		 & \multicolumn{2}{c}{$\Gamma_{\eta\to\gamma e^+ e^-}$} &  \multicolumn{2}{c}{$\text{BR}^{\text{rel}}_{\eta\to\gamma e^+e^-}$} &  \multicolumn{2}{c}{$\Gamma_{\eta'\to\gamma e^+ e^-}$} &  \multicolumn{2}{c}{$\text{BR}^{\text{rel}}_{\eta'\to\gamma e^+e^-}$} \\ 
		& \multicolumn{2}{c}{[$10^{-2}$ eV]} &  \multicolumn{2}{c}{$[10^{-2}]$} & \multicolumn{2}{c}{[eV]} &  \multicolumn{2}{c}{$[10^{-2}]$} &  \multicolumn{2}{c}{ [eV]} &  \multicolumn{2}{c}{$[10^{-2}]$}\\
		\myMR
  \text{LO} & 9.29 & 0.02 & 1.19 & 0. & 10.1 & 0.02 & 1.63 & 0.00 & 90.59 & 0.20 & 1.80 & 0.01 \\
  \text{LO, Qexp} & 0 & 0 & \multicolumn{2}{c}{-} & 3.22 & 0.01 & 1.61 & 0.00 & 150.06 & 0.33 & 1.80 &
  0.01 \\
  \text{NLO 1} & 9.13 & 0.19 & 1.20 & 0.04 & 10.02 & 1.46 & 1.67 & 0.79 & 85.05 & 18.52 & 2.03 & 4.05 \\
  \text{NLO 2} & 8.35 & 0.14 & 1.20 & 0.26 & 8.35 & 1.05 & 1.67 & 3.11 & 73.29 & 15.74 & 1.69 & 2.86 \\
  \text{NLO 3} & 8.46 & 0.45 & 1.19 & 0.21 & 7.22 & 3.02 & 1.68 & 2.78 & 80.56 & 50.58 & 1.70 & 2.26 \\
  \text{NLO, Qexp} & 9.29 & 0.02 & 1.19 & 0.00 & 8.43 & 8.19 & 1.62 & 22.79 & 74.25 & 111.89 & 1.7 & 22.29 \\
  \text{Full} & 9.11 & 0.19 & 1.19 & 0.04 & 8.67 & 0.33 & 1.68 & 0.09 & 85.54 & 4.98 & 1.96 & 0.13 \\
  \text{$C_P=0$} & 9.10 & 0.19 & 1.19 & 0.04 & 8.67 & 0.33 & 1.68 & 0.09 & 81.44 & 4.67 & 1.87 & 0.12 \\
\text{Exp. \cite{Tanabashi:2018oca}}&9.18 &0.33 & 1.19&0.04 &9.04&0.63& 1.75&0.10 & 92.71 & 7.26 & 2.13 & 0.16\\
\myBR	
		\end{tabular}
		\caption{Decay widths and relative BRs for $P\to\gamma e^+e^-$.}
		\label{tab:ee}
		\end{table}

\begin{table}[tbp]
	\centering
		\begin{tabular}{c  r@{$\,\pm\,$}l r@{$\,\pm\,$}l r@{$\,\pm\,$}l r@{$\,\pm\,$}l }\myTR
		 & \multicolumn{2}{c}{$\Gamma_{\eta\to\gamma \mu^+\mu^-}$} &  \multicolumn{2}{c}{$\text{BR}^{\text{rel}}_{\eta\to\gamma \mu^+\mu^-}$} &  \multicolumn{2}{c}{$\Gamma_{\eta'\to\gamma \mu^+\mu^-}$} &  \multicolumn{2}{c}{$\text{BR}^{\text{rel}}_{\eta'\to\gamma \mu^+\mu^-}$} \\ 
			& \multicolumn{2}{c}{[eV]} &  \multicolumn{2}{c}{$[10^{-4}]$} &  \multicolumn{2}{c}{ [eV]} &  \multicolumn{2}{c}{$[10^{-3}]$}\\
			\myMR
\text{LO} & 0.34 & 0.00 & 0.55 & 0.00 & 8.65 & 0.02 & 1.72 & 0.01 \\
\text{LO, Qexp} & 0.11 & 0.00 & 0.54 & 0.00 & 14.32 & 0.03 & 1.72 & 0.01 \\
\text{NLO 1} & 0.44 & 0.06 & 0.73 & 0.34 & 14.71 & 2.67 & 3.50 & 7.00 \\
\text{NLO 2} & 0.35 & 0.06 & 0.70 & 1.31 & 4.02 & 1.76 & 0.93 & 1.61 \\
\text{NLO 3} & 0.31 & 0.12 & 0.71 & 1.17 & 4.57 & 3.83 & 0.96 & 1.39 \\
\text{NLO, Qexp} & 0.30 & 0.29 & 0.58 & 8.17 & 4.42 & 10.54 & 1.01 & 13.41 \\
 \text{Full} & 0.42 & 0.02 & 8.16 & 0.48 & 13.36 & 1.43 & 3.06 & 0.34 \\
\text{$C_P=0$} & 0.42 & 0.01 & 7.96& 0.34 & 9.91 & 0.63 & 2.27 & 0.16 \\
\text{Exp. \cite{Tanabashi:2018oca}}&$ $0.41$$&$$0.05$ $&$ 7.87$&$ 1.02$&$ $21.36$$&$$5.38$ $&$4.91 $&$ 1.23$\\
\myBR	
		\end{tabular}
		\caption{Decay widths and relative BRs for $\eta^{(')}\to\gamma \mu^+\mu^-$.}
		\label{tab:mumu}
		\end{table}

The values for $\Gamma_{P\to\gamma e^+e^-}$, $P=\pi^0,\ \eta,\ \eta'$, behave like the corresponding values for the decays to two real photons. The disagreement of the two-photon-decay widths in some scenarios and the experimental data is reflected in the values for $\Gamma_{P\to\gamma e^+e^-}$ as well. Therefore, we calculate the relative branching ratios (BR)
\begin{align}
\text{BR}^{\text{rel}}_{P\to\gamma l^+l^-}=\frac{\Gamma_{P\to\gamma l^+l^-}}{\Gamma_{P\to\gamma\gamma}},
\end{align}
using the values for $\Gamma_{P\to\gamma\gamma}$ obtained in the different scenarios. The results are shown in Tables \ref{tab:ee} and \ref{tab:mumu}.
Now, the values for the relative BRs do not vary very much within the different cases and orders. The $\pi^0$ relative BRs for the decay to an $e^+e^-$ pair agree with the experimental value and also the $\eta$ relative BR is very close to the experimental one, 
while the $\eta'$ relative BR is somewhat smaller than the experimental result, especially in most of the NLO cases. This is related to the value of the $\eta'$ slope. The slope is very large in the NLO 1 case, which leads to a large relative BR, and the negative values for $b_{\eta'}$ in the other NLO cases are reflected in a reduced relative BR even when compared to the LO value.
The decay width of $P\to\gamma e^+e^-$ receives its main contribution at values where the virtual photon is in the vicinity of its mass shell. 
Therefore, the relative BRs are well described already at LO. 
The decay $P\to\gamma\mu^+\mu^-$ provides a better probe of the virtual behavior of the TFF at larger photon virtualities. However, the values for $\Gamma_{\eta\to\gamma \mu^+\mu^-}$ are still related to the two-photon-decay widths, but the higher-order corrections in $q^2$, parameterized by slope and curvature, become important. Here, we calculate the relative BRs as well. The LO relative BR of the $\eta$ is now lower than the experimental value and increases at NLO and NNLO. Especially at NNLO, we obtain a very good agreement with the data for both $\Gamma_{\eta\to\gamma \mu^+\mu^-}$ and $\text{BR}^{\text{rel}}_{\eta\to\gamma \mu^+\mu^-}$. The LO relative BR for $\eta'\to\gamma \mu^+\mu^-$ is only 30\% of the experimental value. In the NLO scenarios, it becomes even smaller except for NLO 1. This is related to the slope of the $\eta'$, which is very large in the NLO 1 scenario, but poorly described in the other NLO cases, even cases with negative values. The full NNLO value is larger than the LO one and most of the NLO values. However, it is still smaller than the experimental result.
If we neglect the $C_{\eta'}$ term, the relative BR decreases again. This is connected to the description of the $\eta'$ TFF data. The time-like TFF is underestimated for higher values of $q^2$ and even more so if one does not take the ${(q^2)}^2$ term into account. The decay width of $\eta'\to\gamma \mu^+\mu^-$ receives contributions in $q^2$ ranges where vector-meson resonances become important \cite{Escribano:2015vjz}, which are not included in our framework. 

Our full NNLO results for the relative BRs are compared with other theoretical determinations. 
The comparison for the $\pi^0$, $\eta$, and $\eta'$ relative BRs can be found in Tables \ref{tab:CompBRpi0} - \ref{tab:CompBRp}.

\begin{table}[tbp]
	\centering
	\begin{tabular}{c  r@{$\,\pm\,$}l}\myTR
		& \multicolumn{2}{c}{$\text{BR}^{\text{rel}}_{\eta\to\gamma e^+e^-}$ $[10^{-2}]$} \\ \myMR
\text{QED \cite{Miyazaki:1974qi}} & 1.18 & 0 \\
\text{Hidden gauge \cite{Petri:2010ea}} & 1.19 & 0 \\
\text{Mod. VMD \cite{Petri:2010ea}} & 1.19 & 0 \\
\text{Quark model \cite{Lih:2009np}} & 1.18 & 0 \\
\text{$\chi $PT+VM \cite{Terschlusen:2013iqa}} & 1.21 & 0 \\
\text{DA \cite{Hoferichter:2014vra}} & 1.19 & 0 \\
\text{DS Eq. \cite{Weil:2017knt}} & 1.19 & 0.03 \\
\text{Pade approx. \cite{Escribano:2015vjz}} & 1.19 & 0.01 \\
\text{R$\chi $PT \cite{Kimura:2016xnx}} & 1.19 & 0.03 \\
\text{This work} & 1.19 & 0.04 \\
\text{Exp. \cite{Tanabashi:2018oca}} & 1.19 & 0.04 \\
		\myBR
	\end{tabular}
	\caption{Comparison of theoretical determinations of the $\pi^0$ relative BR.}
	\label{tab:CompBRpi0}
\end{table}

\begin{table}[tbp]
	\centering
		\begin{tabular}{c  r@{$\,\pm\,$}l r@{$\,\pm\,$}l }\myTR
		 & \multicolumn{2}{c}{$\text{BR}^{\text{rel}}_{\eta\to\gamma e^+e^-}$ $[10^{-2}]$} &  \multicolumn{2}{c}{$\text{BR}^{\text{rel}}_{\eta\to\gamma \mu^+\mu^-}$ $[10^{-4}]$} \\ \myMR
\text{QED \cite{Miyazaki:1974qi}} & $1.63$&$0$ & $5.54$&$0$ \\
\text{Hidden gauge \cite{Petri:2010ea}} & $1.666$&$0.002$ &
$7.75$&$0.09$ \\
\text{Mod. VMD \cite{Petri:2010ea}} & $1.662$&$0.002$ &
$7.54$&$0.11$ \\
\text{Quark model \cite{Lih:2009np}} & $1.77$&$0$ & $7.48$&$0$
\\
\text{Pade approx. \cite{Escribano:2015vjz}} & $1.68$&$0.15$ &
$8.30$&$1.42$ \\
\text{R$\chi $PT \cite{Kimura:2016xnx}} & $1.66$&$0.06$ &
$7.18$&$0.63$ \\
\text{This work} & $1.68$&$0.09$ & $8.16$&$0.48$ \\
\text{Exp. \cite{Tanabashi:2018oca}}& $ 1.75$&$0.10 $&$ 7.87$&$ 1.02$\\
\myBR
		\end{tabular}
		\caption{Comparison of theoretical determinations of the $\eta$ relative BRs.}
		\label{tab:CompBR}
		\end{table}

\begin{table}[tbp]
	\centering
		\begin{tabular}{c  r@{$\,\pm\,$}l r@{$\,\pm\,$}l }\myTR
		 & \multicolumn{2}{c}{$\text{BR}^{\text{rel}}_{\eta'\to\gamma e^+e^-}$ $[10^{-2}]$} &  \multicolumn{2}{c}{$\text{BR}^{\text{rel}}_{\eta'\to\gamma \mu^+\mu^-}$ $[10^{-3}]$} \\  \myMR
\text{Hidden gauge \cite{Petri:2010ea}} & $2.10$&$0.02$ &
$4.45$&$0.15$ \\
\text{Mod. VMD \cite{Petri:2010ea}} & $2.06$&$0.02$ &
$4.11$&$0.18$ \\
\text{Pade approx. \cite{Escribano:2015vjz}} & $1.99$&$0.16$ &
$3.36$&$0.26$ \\
\text{R$\chi $PT \cite{Kimura:2016xnx}} & $2.00$&$0.13$ &$3.67$&$ 1.15$ \\
\text{This work} & $1.96$&$0.13$ & $3.06$&$0.34$ \\
\text{Exp. \cite{Tanabashi:2018oca}}&$ 2.13 $&$ 0.16$ &$4.91 $&$ 1.23$\\
\myBR	
		\end{tabular}
		\caption{Comparison of theoretical determinations of the $\eta'$ relative BRs.}
		\label{tab:CompBRp}
		\end{table}
		
Our values for $\text{BR}^{\text{rel}}_{P\to\gamma e^+e^-}$, $P=\pi^0,\ \eta,\ \eta'$, agree well with the other determinations. In the case of $\text{BR}^{\text{rel}}_{\eta\to\gamma \mu^+\mu^-}$, as already stated, the simple QED prediction is too small. Here, our result agrees with the other works, except for Ref.~\cite{Lih:2009np} which gives a slightly smaller value. 
Our result for $\text{BR}^{\text{rel}}_{\eta'\to\gamma \mu^+\mu^-}$ is smaller than the others. It agrees within errors with Ref.~\cite{Escribano:2015vjz}, and the determinations including vector mesons are larger.

\section{Summary and outlook}

We have studied the $P\to\gamma^{(\ast)}\gamma^{(\ast)}$ interaction, where $P=\pi^0,\ \eta,\ \eta'$, at the one-loop level up to and including NNLO in L$N_c$ChPT. Besides the loop corrections, all contact terms appearing at NNLO have been calculated, except for those of the $\mathcal{O}(p^8)$ Lagrangian, which has not been constructed yet. However, in the expressions for the form factors describing the decays, possible structures originating from the $\mathcal{O}(p^8)$ Lagrangian have been introduced phenomenologically, accompanied by free parameters. Furthermore, the $\eta$-$\eta'$ mixing at NNLO has been consistently included. The numerical analyses of the decays have been performed successively at LO, NLO, and NNLO. At NLO, we employed the values for the LECs and mixing angle determined in the NLO analysis of the $\eta$-$\eta'$ mixing in Ref.~\cite{Bickert:2016fgy} labeled NLO I. The LECs from the odd-intrinsic-parity sector were fixed to the experimental data of the decay widths to real photons and the slope parameters of $\pi^0$, $\eta$, $\eta'$. We have found that the NLO results are not sufficient to describe all data simultaneously. If the $1/N_c$ expansion of the quark-charge matrix is taken into account, the results worsen. At NNLO, the LECs have been determined through a fit to the experimental data for the $\pi^0$, $\eta$, and $\eta'$ transition form factors. We have achieved a good description of the $\pi^0$ TFF between $-0.45\ \text{GeV}^2$ and $0.45\ \text{GeV}^2$, of the $\eta$ TFF up to $0.45\ \text{GeV}$, and of the $\eta'$ TFF between $-0.25\ \text{GeV}^2$ and $0.3\ \text{GeV}^2$, which is mainly caused by the inclusion of ${(q^2)}^2$ terms, whereas loops do not play an important role. In addition, we have calculated the slopes and the curvatures of the TFFs and the decay widths of $P\to\gamma l^+l^-$, where $l=e,\ \mu$, and compared them to other works. In general, our NNLO results for those quantities tend to agree with the other experimental and theoretical determinations.

Our results clearly indicate that a perturbative chiral and 1/$N_c$ expansion has its limitations in the $\pi^0$-$\eta$-$\eta'$ system. While going to even higher orders in the expansion might result in an improved description of experimental data, this would involve additional unknown LECs, making the gain in physical insight questionable. However, with reference to the transition form factors, an extended theory including vector-meson degrees of freedom might improve the situation with respect to larger values of $|q^2|$, in particular, in the time-like region. In addition to a purely phenomenological treatment, one might set up a power-counting scheme in terms of the complex-mass renormalization \cite{Djukanovic:2009zn} like in the calculation of the vector form factor of the pion \cite{Djukanovic:2014rua}.

\begin{acknowledgements}
Supported by the Deutsche Forschungsgemeinschaft DFG through the Collaborative Research Center “The Low-Energy Frontier of the Standard Model” (SFB 1044). We would like to thank A.~Denig for useful comments on the experimental results.
\end{acknowledgements}

\begin{appendix}

\section{Additional expressions}\label{app:expAnomD}

In the case without the $1/N_c$ expansion of the quark-charge matrix $Q$, the loop contributions to the form factors of the two-photon decays, given by the loop diagrams in Fig.~\ref{fig:phiggallp}, read
\begin{align}
	&F_{\pi^0\gamma^*\gamma^*}\no\\
	&=\frac{1}{1152 \pi ^4 F_{\pi
		}^3}\left\{3 \left(q_1^2-4 M_K^2\right) B_0\left(q_1^2,M_K^2,M_K^2\right)\right.\no\\
	&\quad+3 \left(q_2^2-4
	M_K^2\right) B_0\left(q_2^2,M_K^2,M_K^2\right)+3 \left[\left(q_1^2-4 M_{\pi }^2\right)
	B_0\left(q_1^2,M_{\pi }^2,M_{\pi }^2\right)\right.\no\\
	&\quad\left.+\left(q_2^2-4 M_{\pi }^2\right)
	B_0\left(q_2^2,M_{\pi }^2,M_{\pi }^2\right)\right]+24 A_0\left(M_K^2\right)+24 A_0\left(M_{\pi
	}^2\right)\no\\
	&\quad\left.+4 \left[-6 \left(M_K^2+M_{\pi }^2\right)+q_1^2+q_2^2\right]\right\},
\end{align}
\begin{align}
	&F_{\eta\gamma^*\gamma^*}\no\\
	&=\frac{1}{1152 \sqrt{3} \pi ^4 F_{\pi }^3}\left(\left[\sqrt{2} \sin (\theta^{[0]} )-\cos (\theta^{[0]} )\right] \left\{-q_2^2\left[3
	B_0\left(q_2^2,M_K^2,M_K^2\right)\right.\right.\right.\no\\
	&\quad\left.+3 B_0\left(q_2^2,M_{\pi }^2,M_{\pi
	}^2\right)+4\right]+3 \left(4 M_K^2-q_1^2\right) B_0\left(q_1^2,M_K^2,M_K^2\right)\no\\
	&\quad+4 
	\left[3 M_K^2 B_0\left(q_2^2,M_K^2,M_K^2\right)+3 M_{\pi }^2 B_0\left(q_2^2,M_{\pi }^2,M_{\pi
	}^2\right)+6 \left(M_K^2+M_{\pi }^2\right)-q_1^2\right]\no\\
	&\quad\left.+3 \left(4 M_{\pi }^2-q_1^2\right)
	B_0\left(q_1^2,M_{\pi }^2,M_{\pi }^2\right)\right\}+6 \left[\cos (\theta^{[0]} )-7 \sqrt{2} \sin
	(\theta^{[0]} )\right] A_0\left(M_K^2\right)\no\\
	&\quad\left.+6 A_0\left(M_{\pi }^2\right) \left[7 \cos (\theta^{[0]} )-10
	\sqrt{2} \sin (\theta^{[0]} )\right]\right),
\end{align}
\begin{align}
	&F_{\eta'\gamma^*\gamma^*}\no\\
	&=\frac{1}{1152 \sqrt{3} \pi ^4 F_{\pi }^3}\left(-\left[\sin (\theta^{[0]} )+\sqrt{2} \cos (\theta^{[0]} )\right] \left\{-q_2^2\left[3
	B_0\left(q_2^2,M_K^2,M_K^2\right)\right.\right.\right.\no\\
	&\quad\left.+3 B_0\left(q_2^2,M_{\pi }^2,M_{\pi
	}^2\right)+4\right]+3 \left(4 M_K^2-q_1^2\right) B_0\left(q_1^2,M_K^2,M_K^2\right)\no\\
	&\quad+4
	\left[3 M_K^2 B_0\left(q_2^2,M_K^2,M_K^2\right)+3 M_{\pi }^2 B_0\left(q_2^2,M_{\pi }^2,M_{\pi
	}^2\right)+6 \left(M_K^2+M_{\pi }^2\right)-q_1^2\right]\no\\
	&\quad\left.+3 \left(4 M_{\pi }^2-q_1^2\right)
	B_0\left(q_1^2,M_{\pi }^2,M_{\pi }^2\right)\right\}+6 \left[\sin (\theta^{[0]} )+7 \sqrt{2} \cos
	(\theta^{[0]} )\right] A_0\left(M_K^2\right)\no\\
	&\quad\left.+6 A_0\left(M_{\pi }^2\right) \left[7 \sin (\theta^{[0]} )+10
	\sqrt{2} \cos (\theta^{[0]} )\right]\right).
\end{align}
Including the $1/N_c$ expansion of $Q$, the loop contributions are given by
\begin{align}
	&F_{\pi^0\gamma^*\gamma^*}\no\\
	&=\frac{1}{2304 \pi ^4 F_{\pi }^3}\left[3 \left(q_1^2-4 M_K^2\right) B_0\left(q_1^2,M_K^2,M_K^2\right)\right.\no\\
	&\quad\left.+3 \left(q_2^2-4
	M_K^2\right) B_0\left(q_2^2,M_K^2,M_K^2\right)+24 A_0\left(M_K^2\right)+2 \left(-12
	M_K^2+q_1^2+q_2^2\right)\right],
\end{align}
\begin{align}
	&F_{\eta\gamma^*\gamma^*}\no\\
	&=\frac{1}{4608 \sqrt{3} \pi ^4 F_{\pi
		}^3}\left(\sqrt{2} \sin (\theta^{[0]} )
	\left\{2 \left[2 \left\{-q_2^2\left[3 B_0\left(q_2^2,M_K^2,M_K^2\right)+3
	B_0\left(q_2^2,M_{\pi }^2,M_{\pi }^2\right)+4\right]\right.\right.\right.\right.\no\\
	&\quad+3 \left(4 M_K^2-q_1^2\right)
	B_0\left(q_1^2,M_K^2,M_K^2\right)+4 \left[3 M_K^2 B_0\left(q_2^2,M_K^2,M_K^2\right)\right.\no\\
	&\quad\left.\left.+3 M_{\pi }^2
	B_0\left(q_2^2,M_{\pi }^2,M_{\pi }^2\right)+6 \left(M_K^2+M_{\pi }^2\right)-q_1^2\right]+3
	\left(4 M_{\pi }^2-q_1^2\right) B_0\left(q_1^2,M_{\pi }^2,M_{\pi }^2\right)\right\}\no\\
	&\quad\left.\left.-129
	A_0\left(M_{\pi }^2\right)\right]-177 A_0\left(M_K^2\right)\right\}+2 \cos (\theta^{[0]} ) \left\{3
	\left(4 M_K^2-q_1^2\right) B_0\left(q_1^2,M_K^2,M_K^2\right)\right.\no\\
	&\quad+3 \left(4 M_K^2-q_2^2\right)
	B_0\left(q_2^2,M_K^2,M_K^2\right)+6 \left[\left(q_1^2-4 M_{\pi }^2\right) B_0\left(q_1^2,M_{\pi
	}^2,M_{\pi }^2\right)\right.\no\\
	&\quad\left.+\left(q_2^2-4 M_{\pi }^2\right) B_0\left(q_2^2,M_{\pi }^2,M_{\pi
	}^2\right)\right]-24 A_0\left(M_K^2\right)+48 A_0\left(M_{\pi }^2\right)\no\\
	&\quad\left.\left.+2 \left[12
	\left(M_K^2-2 M_{\pi }^2\right)+q_1^2+q_2^2\right]\right\}\right),
\end{align}
\begin{align}
	&F_{\eta'\gamma^*\gamma^*}\no\\
	&=\frac{1}{4608 \sqrt{3} \pi ^4 F_{\pi }^3}\left\{2 \sin (\theta^{[0]} ) \left\{3
	\left(4 M_K^2-q_1^2\right) B_0\left(q_1^2,M_K^2,M_K^2\right)\right.\right.\no\\
	&\quad+3 \left(4 M_K^2-q_2^2\right)
	B_0\left(q_2^2,M_K^2,M_K^2\right)+6 \left[\left(q_1^2-4 M_{\pi }^2\right) B_0\left(q_1^2,M_{\pi
	}^2,M_{\pi }^2\right)\right.\no\\
	&\quad\left.+\left(q_2^2-4 M_{\pi }^2\right) B_0\left(q_2^2,M_{\pi }^2,M_{\pi
	}^2\right)\right]-24 A_0\left(M_K^2\right)+48 A_0\left(M_{\pi }^2\right)\no\\
	&\quad\left.+2 \left[12
	\left(M_K^2-2 M_{\pi }^2\right)+q_1^2+q_2^2\right]\right\}+\sqrt{2} \cos (\theta^{[0]} ) \left[4
	\left\{3 \left(q_1^2-4 M_K^2\right) B_0\left(q_1^2,M_K^2,M_K^2\right)\right.\right.\no\\
	&\quad+3 \left(q_2^2-4
	M_K^2\right) B_0\left(q_2^2,M_K^2,M_K^2\right)+3 \left[\left(q_1^2-4 M_{\pi }^2\right)
	B_0\left(q_1^2,M_{\pi }^2,M_{\pi }^2\right)\right.\no\\
	&\quad\left.\left.+\left(q_2^2-4 M_{\pi }^2\right)
	B_0\left(q_2^2,M_{\pi }^2,M_{\pi }^2\right)\right]+4 \left[-6 \left(M_K^2+M_{\pi
	}^2\right)+q_1^2+q_2^2\right]\right\}\no\\
	&\quad\left.\left.+177 A_0\left(M_K^2\right)+258 A_0\left(M_{\pi
	}^2\right)\right]\right\}.
\end{align}
	
\section{Fit parameters}\label{app:FitParameters}

\begin{table}
	\centering
	\begin{tabular}{c c r@{$\,\pm\,$}l r@{$\,\pm\,$}l r@{\,$\pm$\,}l}\myTR
		& $|q^2|_\text{max}$ [GeV$^2$]	 & \multicolumn{2}{c}{$A_{\pi^0}$} & \multicolumn{2}{c}{$B_{\pi^0}\ [\text{GeV}^{-2}]$} &  \multicolumn{2}{c}{$C_{\pi^0}\ [\text{GeV}^{-4}]$}\\ \myMR
		 \text{Full} & 0.55 & $-0.01$&$0.01$ & $1.24$&$0.10$ & $1.77$&$0.18$ \\
		\text{Full} & 0.5 & $-0.01$&$0.01$ & $0.91$&$0.15$ & $1.40$&$0.35$ \\
		\text{Full} & 0.45 & $-0.01$&$0.01$ & $0.81$&$0.19$ & $1.15$&$0.43$ \\
		\text{Full} & 0.4 & $-0.01$&$0.01$ & $1.08$&$0.29$ & $2.04$&$0.83$ \\
		\text{Full, Qexp} & 0.55 & $0.99$&$0.01$ & $1.57$&$0.10$ & $1.95$&$0.18$ \\
		\text{Full, Qexp} & 0.5 & $0.99$&$0.01$ & $1.25$&$0.15$ & $1.60$&$0.35$ \\
		\text{Full, Qexp} & 0.45 & $0.99$&$0.01$ & $1.15$&$0.19$ & $1.36$&$0.43$ \\
		\text{Full, Qexp} & 0.4 & $0.99$&$0.01$ & $1.43$&$0.29$ & $2.29$&$0.83$ \\
		\text{W/o loops} & 0.55 & $-0.01$&$0.01$ & $1.59$&$0.10$ & $1.97$&$0.18$ \\
		\text{W/o loops} & 0.5 & $-0.01$&$0.01$ & $1.28$&$0.15$ & $1.62$&$0.35$ \\
		\text{W/o loops} & 0.45 & $-0.01$&$0.01$ & $1.18$&$0.19$ & $1.39$&$0.43$ \\
		\text{W/o loops} & 0.4 & $-0.01$&$0.01$ & $1.46$&$0.29$ & $2.31$&$0.83$ \\
		\text{W/o loops, Qexp} & 0.55 & $0.99$&$0.01$ & $1.59$&$0.10$ & $1.97$&$0.18$ \\
		\text{W/o loops, Qexp} & 0.5 & $0.99$&$0.01$ & $1.28$&$0.15$ & $1.62$&$0.35$ \\
		\text{W/o loops, Qexp} & 0.45 & $0.99$&$0.01$ & $1.18$&$0.19$ & $1.39$&$0.43$ \\
		\text{W/o loops, Qexp} & 0.4 & $0.99$&$0.01$ & $1.46$&$0.29$ & $2.31$&$0.83$ \\
		$C_{\pi^0}=0$ & 0.55 & $-0.01$&$0.01$ & $0.33$&$0.03$ & $0$&$0$ \\
		$C_{\pi^0}=0$ & 0.5 & $-0.01$&$0.01$ & $0.34$&$0.06$ & $0$&$0$ \\
		$C_{\pi^0}=0$ & 0.45 & $-0.01$&$0.01$ & $0.33$&$0.06$ & $0$&$0$ \\
		$C_{\pi^0}=0$ & 0.4 & $-0.01$&$0.01$ & $0.42$&$0.13$ & $0$&$0$ \\
		$C_{\pi^0}=0$, Qexp & 0.55 & $0.99$&$0.01$ & $0.56$&$0.03$ & $0$&$0$ \\
		$C_{\pi^0}=0$, Qexp & 0.5 & $0.99$&$0.01$ & $0.60$&$0.06$ & $0$&$0$ \\
		$C_{\pi^0}=0$, Qexp & 0.45 & $0.99$&$0.01$ & $0.59$&$0.06$ & $0$&$0$ \\
		$C_{\pi^0}=0$, Qexp & 0.4 & $0.99$&$0.01$ & $0.70$&$0.13$ & $0$&$0$ \\
		\text{W/o loops $\land $ }$C_{\pi^0}=0$ & 0.55 & $-0.01$&$0.01$ & $0.58$&$0.03$ & $0$&$0$ \\
		\text{W/o loops $\land $ }$C_{\pi^0}=0$ & 0.5 & $-0.01$&$0.01$ & $0.61$&$0.06$ & $0$&$0$ \\
		\text{W/o loops $\land $ }$C_{\pi^0}=0$ & 0.45 & $-0.01$&$0.01$ & $0.60$&$0.06$ & $0$&$0$ \\
		\text{W/o loops $\land $ }$C_{\pi^0}=0$ & 0.4 & $-0.01$&$0.01$ & $0.71$&$0.13$ & $0$&$0$ \\
		\text{W/o loops $\land $ }$C_{\pi^0}=0$, Qexp & 0.55 & $0.99$&$0.01$ & $0.58$&$0.03$ & $0$&$0$ \\
		\text{W/o loops $\land $ }$C_{\pi^0}=0$, Qexp & 0.5 & $0.99$&$0.01$ & $0.61$&$0.06$ & $0$&$0$ \\
		\text{W/o loops $\land $ }$C_{\pi^0}=0$, Qexp & 0.45 & $0.99$&$0.01$ & $0.60$&$0.06$ & $0$&$0$ \\
		\text{W/o loops $\land $ }$C_{\pi^0}=0$, Qexp & 0.4 & $0.99$&$0.01$ & $0.71$&$0.13$ & $0$&$0$ \\
		\myBR
	\end{tabular}
	\caption{Fit parameters for the $\pi^0$ TFF.}
	\label{tab:pi0NNLOFull}
\end{table}

\begin{table}
	\centering
	\begin{tabular}{c c r@{$\,\pm\,$}l r@{$\,\pm\,$}l r@{\,$\pm$\,}l}\myTR
		& $m_\text{max}(l^+l^-)$ [GeV] & \multicolumn{2}{c}{$A_\eta$} & \multicolumn{2}{c}{$B_\eta\ [\text{GeV}^{-2}]$} &  \multicolumn{2}{c}{$C_\eta\ [\text{GeV}^{-4}]$}\\ \myMR
		 \text{Full} & $0.47$ & $-0.17$&$0.03$ & $2.32$&$0.22$ & $10.51$&$1.82$ \\
		\text{Full} & $0.40$ & $-0.17$&$0.03$ & $2.58$&$0.26$ & $7.26$&$2.59$ \\
		\text{Full} & $0.35$ & $-0.17$&$0.03$ & $2.67$&$0.35$ & $5.89$&$4.29$ \\
		\text{Full, Qexp} & $0.47$ & $0.66$&$0.03$ & $2.39$&$0.22$ & $10.59$&$1.82$ \\
		\text{Full, Qexp} & $0.40$ & $0.66$&$0.03$ & $2.66$&$0.26$ & $7.34$&$2.59$ \\
		\text{Full, Qexp} & $0.35$ & $0.66$&$0.03$ & $2.74$&$0.35$ & $5.96$&$4.29$ \\
		\text{W/o loops} & $0.47$ & $-0.17$&$0.03$ & $2.98$&$0.22$ & $11.35$&$1.82$ \\
		\text{W/o loops} & $0.40$ & $-0.17$&$0.03$ & $3.20$&$0.26$ & $8.73$&$2.59$ \\
		\text{W/o loops} & $0.35$ & $-0.17$&$0.03$ & $3.23$&$0.35$ & $8.12$&$4.29$ \\
		\text{W/o loops, Qexp} & $0.47$ & $0.66$&$0.03$ & $2.98$&$0.22$ & $11.35$&$1.82$ \\
		\text{W/o loops, Qexp} & $0.40$ & $0.66$&$0.03$ & $3.20$&$0.26$ & $8.73$&$2.59$ \\
		\text{W/o loops, Qexp} & $0.35$ & $0.66$&$0.03$ & $3.23$&$0.35$ & $8.12$&$4.29$ \\
		$C_{\eta}=0$ & $0.47$ & $-0.17$&$0.03$ & $3.41$&$0.12$ & $0$&$0$ \\
		$C_{\eta}=0$ & $0.40$ & $-0.17$&$0.03$ & $3.25$&$0.12$ & $0$&$0$ \\
		$C_{\eta}=0$ & $0.35$ & $-0.17$&$0.03$ & $3.11$&$0.13$ & $0$&$0$ \\
		$C_{\eta}=0$, Qexp & $0.47$ & $0.66$&$0.03$ & $3.49$&$0.12$ & $0$&$0$ \\
		$C_{\eta}=0$, Qexp & $0.40$ & $0.66$&$0.03$ & $3.33$&$0.12$ & $0$&$0$ \\
		$C_{\eta}=0$, Qexp & $0.35$ & $0.66$&$0.03$ & $3.19$&$0.13$ & $0$&$0$ \\
		\text{W/o loops $\land $ }$C_{\eta}=0$ & $0.47$ & $-0.17$&$0.03$ & $4.16$&$0.12$ & $0$&$0$ \\
		\text{W/o loops $\land $ }$C_{\eta}=0$ & $0.40$ & $-0.17$&$0.03$ & $4.00$&$0.12$ & $0$&$0$ \\
		\text{W/o loops $\land $ }$C_{\eta}=0$ & $0.35$ & $-0.17$&$0.03$ & $3.84$&$0.13$ & $0$&$0$ \\
		\text{W/o loops $\land $ }$C_{\eta}=0$, Qexp & $0.47$ & $0.66$&$0.03$ & $4.16$&$0.12$ & $0$&$0$ \\
		\text{W/o loops $\land $ }$C_{\eta}=0$, Qexp & $0.40$ & $0.66$&$0.03$ & $4.00$&$0.12$ & $0$&$0$ \\
		\text{W/o loops $\land $ }$C_{\eta}=0$, Qexp & $0.35$ & $0.66$&$0.03$ & $3.84$&$0.13$ & $0$&$0$ \\
		\myBR
	\end{tabular}
	\caption{Fit parameters for the $\eta$ TFF.}
	\label{tab:etaNNLOFull}
\end{table}

\begin{table}
	\centering
	\begin{tabular}{c c r@{$\,\pm\,$}l r@{$\,\pm\,$}l r@{\,$\pm$\,}l}\myTR
	& Fit	& \multicolumn{2}{c}{$A_{\eta'}$} & \multicolumn{2}{c}{$B_{\eta'}\ [\text{GeV}^{-2}]$} &  \multicolumn{2}{c}{$C_{\eta'}\ [\text{GeV}^{-4}]$}\\ \myMR
	 \text{Full} & \text{I} & $-0.06$&$0.02$ & $1.08$&$0.25$ & $1.18$&$0.52$ \\
	\text{Full} & \text{II} & $-0.06$&$0.02$ & $0.95$&$0.27$ & $0.92$&$0.55$ \\
	\text{Full} & \text{III} & $-0.06$&$0.02$ & $1.12$&$0.23$ & $2.89$&$1.05$ \\
	\text{Full} & \text{IV} & $-0.06$&$0.02$ & $1.04$&$0.26$ & $2.43$&$1.23$ \\
	\text{Full, Qexp} & \text{I} & $-0.29$&$0.02$ & $1.07$&$0.25$ & $1.17$&$0.52$ \\
	\text{Full, Qexp} & \text{II} & $-0.29$&$0.02$ & $0.95$&$0.27$ & $0.91$&$0.55$ \\
	\text{Full, Qexp} & \text{III} & $-0.29$&$0.02$ & $1.11$&$0.23$ & $2.89$&$1.05$ \\
	\text{Full, Qexp} & \text{IV} & $-0.29$&$0.02$ & $1.03$&$0.26$ & $2.42$&$1.23$ \\
	\text{W/o loops} & \text{I} & $-0.06$&$0.02$ & $1.23$&$0.26$ & $1.30$&$0.52$ \\
	\text{W/o loops} & \text{II} & $-0.06$&$0.02$ & $1.10$&$0.27$ & $1.04$&$0.56$ \\
	\text{W/o loops} & \text{III} & $-0.06$&$0.02$ & $1.27$&$0.23$ & $3.03$&$1.05$ \\
	\text{W/o loops} & \text{IV} & $-0.06$&$0.02$ & $1.19$&$0.26$ & $2.59$&$1.23$ \\
	\text{W/o loops, Qexp} & \text{I} & $-0.29$&$0.02$ & $1.23$&$0.26$ & $1.30$&$0.52$ \\
	\text{W/o loops, Qexp} & \text{II} & $-0.29$&$0.02$ & $1.10$&$0.27$ & $1.04$&$0.56$ \\
	\text{W/o loops, Qexp} & \text{III} & $-0.29$&$0.02$ & $1.27$&$0.23$ & $3.03$&$1.05$ \\
	\text{W/o loops, Qexp} & \text{IV} & $-0.29$&$0.02$ & $1.19$&$0.26$ & $2.59$&$1.23$ \\
	\text{$C_{\eta'}=0$} & \text{I} & $-0.06$&$0.02$ & $0.55$&$0.10$ & $0$&$0$ \\
	\text{$C_{\eta'}=0$} & \text{II} & $-0.06$&$0.02$ & $0.54$&$0.10$ & $0$&$0$ \\
	\text{$C_{\eta'}=0$} & \text{III} & $-0.06$&$0.02$ & $0.78$&$0.23$ & $0$&$0$ \\
	\text{$C_{\eta'}=0$} & \text{IV} & $-0.06$&$0.02$ & $0.72$&$0.23$ & $0$&$0$ \\
	\text{$C_{\eta'}=0$, Qexp} & \text{I} & $-0.29$&$0.02$ & $0.54$&$0.10$ & $0$&$0$ \\
	\text{$C_{\eta'}=0$, Qexp} & \text{II} & $-0.29$&$0.02$ & $0.53$&$0.10$ & $0$&$0$ \\
	\text{$C_{\eta'}=0$, Qexp} & \text{III} & $-0.29$&$0.02$ & $0.77$&$0.23$ & $0$&$0$ \\
	\text{$C_{\eta'}=0$, Qexp} & \text{IV} & $-0.29$&$0.02$ & $0.71$&$0.23$ & $0$&$0$ \\
	\text{W/o loops $\wedge$ $C_{\eta'}=0$} & \text{I} & $-0.06$&$0.02$ & $0.64$&$0.11$ & $0$&$0$ \\
	\text{W/o loops $\wedge$ $C_{\eta'}=0$} & \text{II} & $-0.06$&$0.02$ & $0.63$&$0.10$ & $0$&$0$ \\
	\text{W/o loops $\wedge$ $C_{\eta'}=0$} & \text{III} & $-0.06$&$0.02$ & $0.91$&$0.23$ & $0$&$0$ \\
	\text{W/o loops $\wedge$ $C_{\eta'}=0$} & \text{IV} & $-0.06$&$0.02$ & $0.85$&$0.23$ & $0$&$0$ \\
	\text{W/o loops $\wedge$ $C_{\eta'}=0$, Qexp} & \text{I} & $-0.29$&$0.02$ & $0.64$&$0.11$ & $0$&$0$ \\
	\text{W/o loops $\wedge$ $C_{\eta'}=0$, Qexp} & \text{II} & $-0.29$&$0.02$ & $0.63$&$0.10$ & $0$&$0$ \\
	\text{W/o loops $\wedge$ $C_{\eta'}=0$, Qexp} & \text{III} & $-0.29$&$0.02$ & $0.91$&$0.23$ & $0$&$0$ \\
	\text{W/o loops $\wedge$ $C_{\eta'}=0$, Qexp} & \text{IV} & $-0.29$&$0.02$ & $0.85$&$0.23$ & $0$&$0$ \\
		\myBR
	\end{tabular}
	\caption{Fit parameters for the $\eta'$ TFF. The fit ranges are: $-0.53\ \text{GeV}^2\leq q^2 \leq 0.43\ \text{GeV}^2$ (I), $-0.53\ \text{GeV}^2\leq q^2 \leq 0.40\ \text{GeV}^2$ (II), $-0.50\ \text{GeV}^2\leq q^2 \leq 0.43\ \text{GeV}^2$ (III), $-0.50\ \text{GeV}^2\leq q^2 \leq 0.40\ \text{GeV}^2$
		(IV).}
	\label{tab:etapNNLOFull}
\end{table}

\section{Additional plots}\label{app:plots}

\begin{figure}[htbp]
	\centering
	\includegraphics[width=0.75\textwidth]{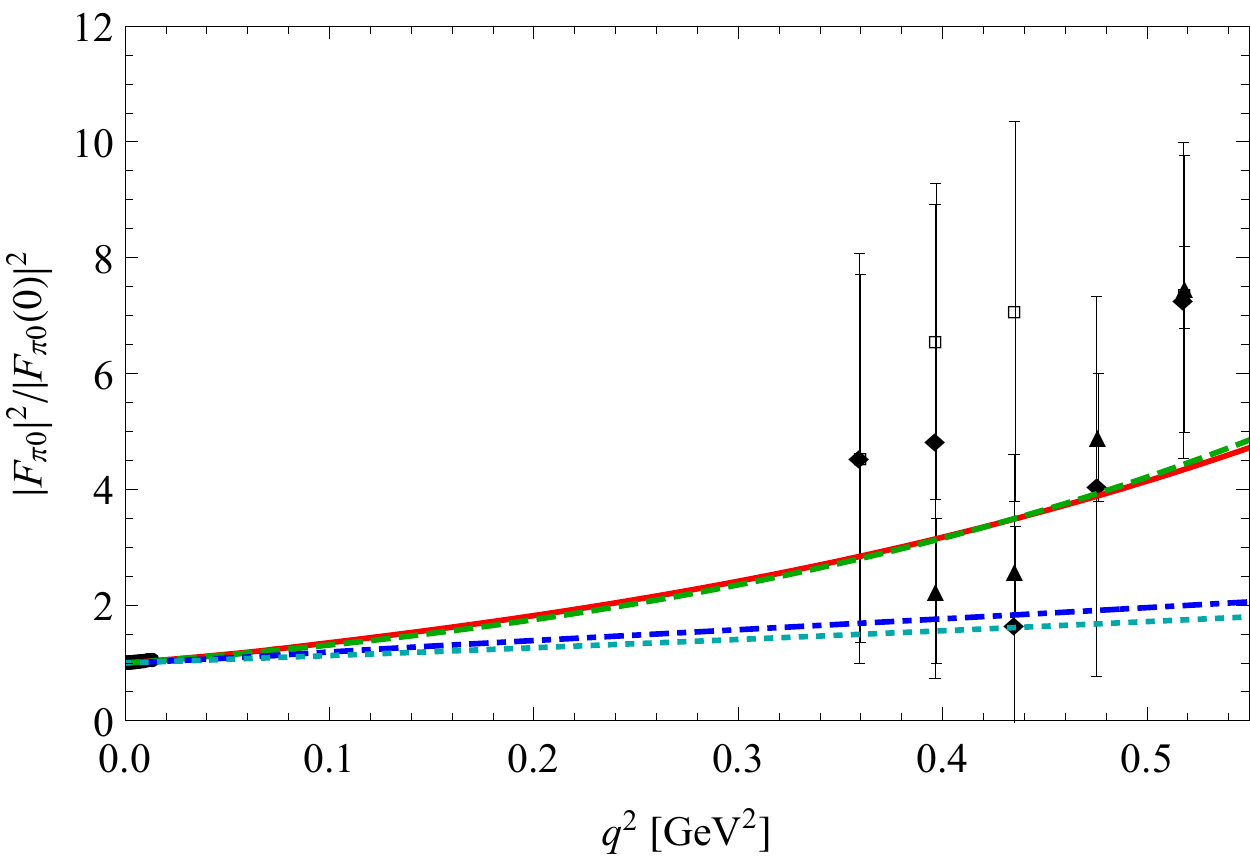}
	\caption{$\pi^0$ TFF in the time-like region, fitted in the range $-0.5\ \text{GeV}^2\leq q^2 \leq 0.5\ \text{GeV}^2$. The solid (red) line is the full NNLO calculation, the dashed (green) line the NNLO result without loops. The blue lines are the NNLO results with $C_{\pi^0}=0$ including loops (dark blue, dash-dotted) and without loops (light blue, dotted). The experimental data are taken from Refs.~\cite{Achasov:2003ed} ($\blacklozenge$), \cite{Akhmetshin:2004gw} ($\square$), \cite{Achasov:2016bfr} ($\blacktriangle$), \cite{Adlarson:2016hpp} ($\bullet$).}
	\label{fig:TFFcasespi0Tl}
\end{figure}

\begin{figure}[htbp]
	\centering
	\includegraphics[width=0.75\textwidth]{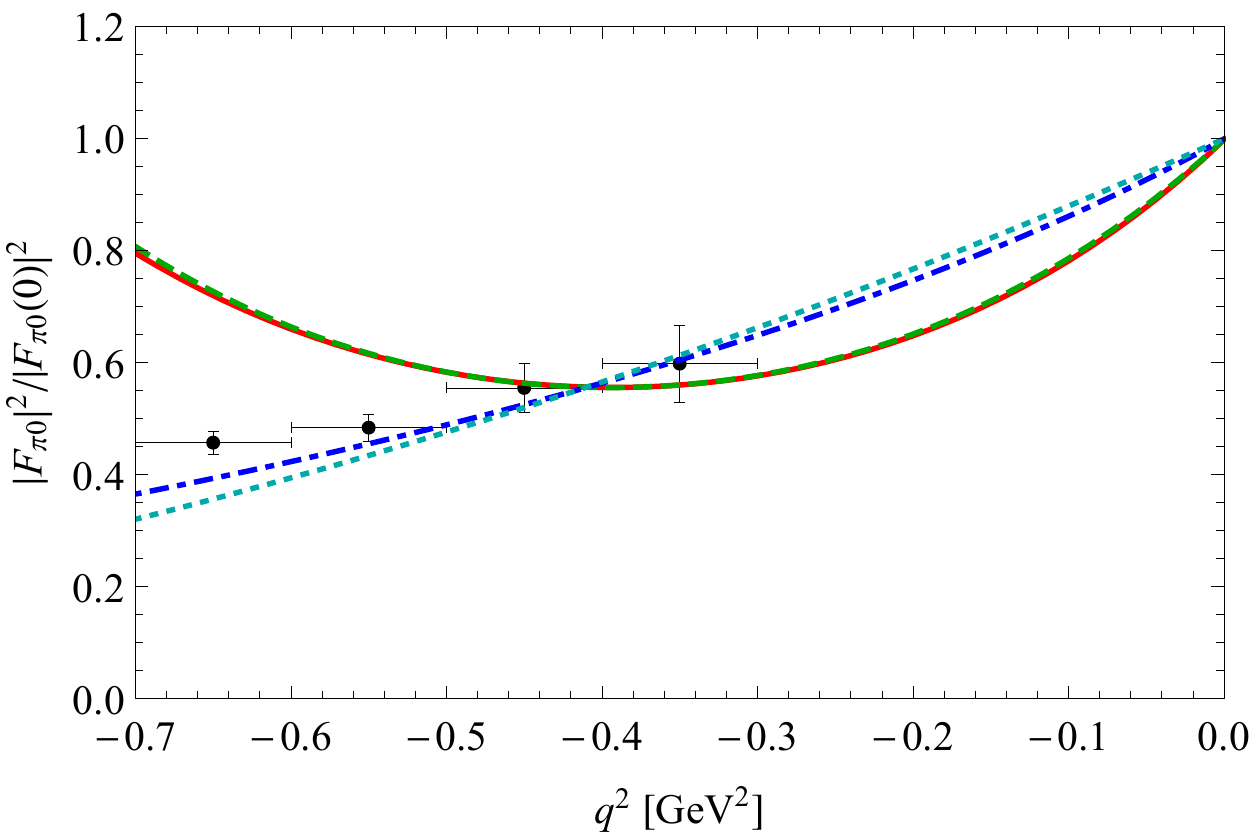}
	\caption{$\pi^0$ TFF in the space-like region, fitted in the range $-0.5\ \text{GeV}^2\leq q^2 \leq 0.5\ \text{GeV}^2$. The solid (red) line is the full NNLO calculation, the dashed (green) line the NNLO result without loops. The blue lines are the NNLO results with $C_{\pi^0}=0$ including loops (dark blue, dash-dotted) and without loops (light blue, dotted). The experimental
		data are taken from Ref.~\cite{Danilkin:2019mhd}.}
	\label{fig:TFFcasespi0Sl}
\end{figure}

\begin{figure}[htbp]
	\centering
	\includegraphics[width=0.9\textwidth]{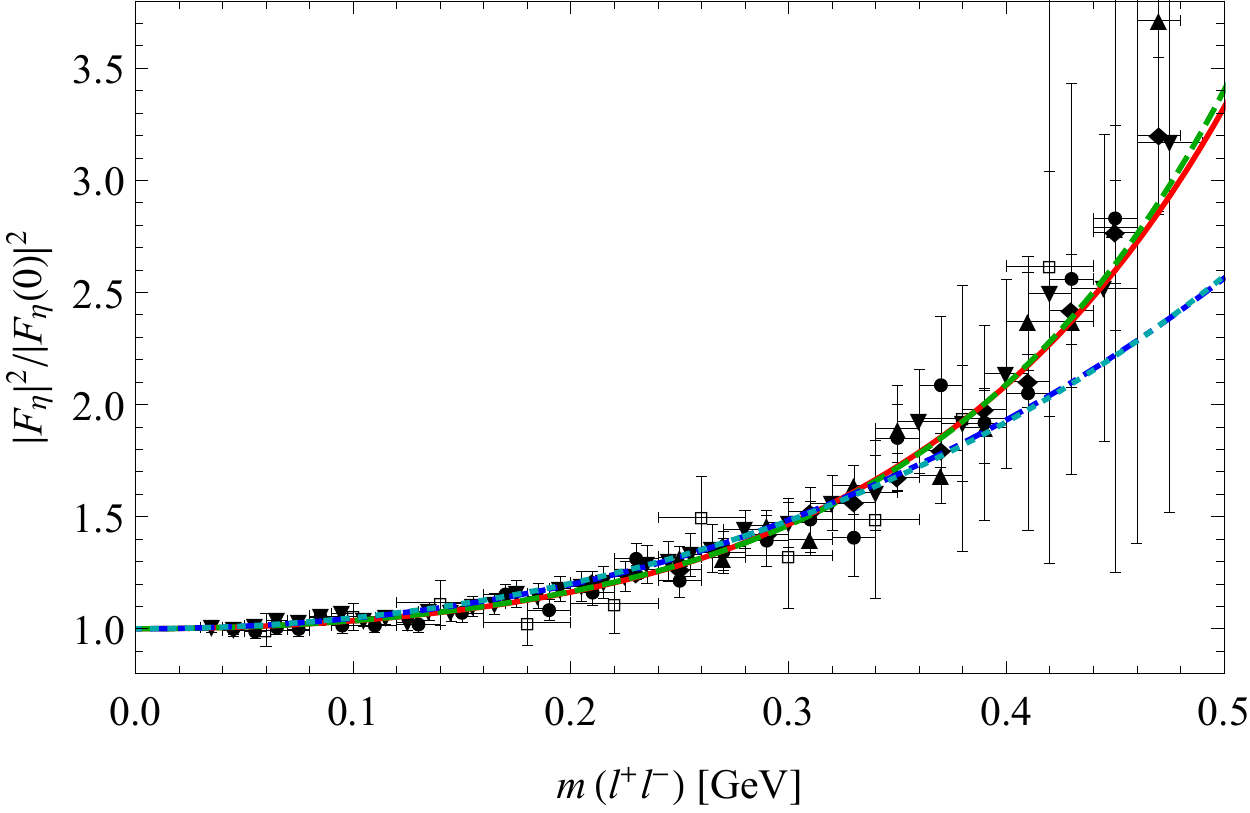}
	\caption{$\eta$ TFF fitted up to $0.47$ GeV. The solid (red) line is the full NNLO calculation, the dashed (green) line the NNLO result without loops. The blue lines are the NNLO results with $C_{\eta}=0$ including loops (dark blue, dash-dotted) and without loops (light blue, dotted). The experimental data are taken from Refs.~\cite{Arnaldi:2009aa} ($\blacktriangle$), \cite{Berghauser:2011zz} ($\square$), \cite{Aguar-Bartolome:2013vpw} ($\blacksquare$), \cite{Arnaldi:2016pzu} ($\blacklozenge$).}
	\label{fig:TFFcases}
\end{figure}

\begin{figure}[htbp]
	\centering
	\includegraphics[width=0.9\textwidth]{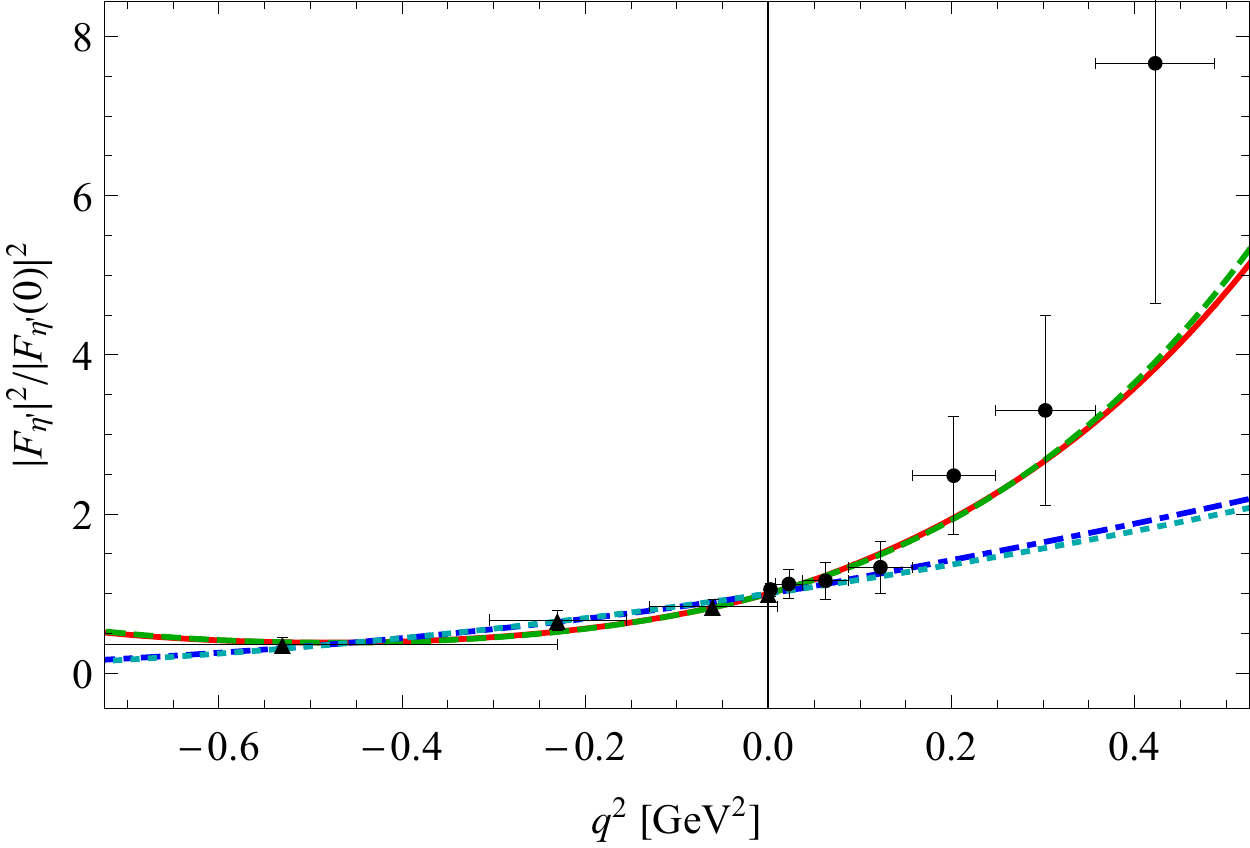}
	\caption{$\eta'$ TFF fitted between $-0.53\ \text{GeV}^2$ and $0.43\ \text{GeV}^2$. The solid (red) line is the full NNLO calculation, the dashed (green) line the NNLO result without loops. The blue lines are the NNLO results with $C_{\eta'}=0$ including loops (dark blue, dash-dotted) and without loops (light blue, dotted). The time-like data are taken from Ref.~\cite{Ablikim:2015wnx} ($\bullet$) and the space-like data from Ref.~\cite{Acciarri:1997yx} ($\blacktriangle$).}
	\label{fig:TFFcasesp}
\end{figure}
	
\end{appendix}

\end{document}